\documentclass[a4paper,11pt]{article}
\pdfoutput=1 

\usepackage{jheppub} 

\usepackage[T1]{fontenc} 
\usepackage[utf8]{inputenc}
\usepackage[dvipsnames]{xcolor}
\usepackage{slashed}
\usepackage[normalem]{ulem}
\usepackage{sidecap}
\usepackage{mathtools}
\usepackage{caption}
\usepackage{subcaption}
\usepackage{multicol}
\usepackage[toc]{appendix}

\newcommand{\Varc}{\langle \tilde{\chi}^2 \rangle}
\newcommand{\CL}{{\tt ${\mathcal C}$osmo${\mathcal L}$attice}}
\def\figureautorefname~#1\null{Fig.\,#1\null}
\def\tableautorefname~#1\null{Tab.\,#1\null}

\def\equationautorefname~#1\null{Eq.\,(#1)\null}

\title{Gravitational wave production from 
preheating with trilinear interactions}
\author{Catarina Cosme,}
\author{Daniel G. Figueroa,}
\author{and Nicol\'{a}s Loayza}
\affiliation{Instituto de F\'{i}sica Corpuscular (IFIC), Universitat de Val\`{e}ncia-CSIC, Parc Cient\'{i}fic UV, C/ Catedr\'{a}tico Jos\'{e} Beltr\'{a}n 2,
E-46980 Paterna, Spain}

\emailAdd{catarina.cosme@ific.uv.es}
\emailAdd{daniel.figueroa@ific.uv.es}
\emailAdd{nicolas.loayza@ific.uv.es}

\abstract{We investigate the production of gravitational waves (GWs) during preheating with monomial/polynomial inflationary potentials, 
considering a trilinear coupling $\phi\chi^2$ between a daughter field $\chi$ and the inflaton $\phi$. For sufficiently large couplings, the trilinear interaction leads to an exponential production of $\chi$ particles, and as a result, a large stochastic GW background (SGWB) is generated throughout the process. We study the linear and non-linear dynamics of preheating with lattice simulations, following the production of GWs through all relevant stages. We find that large couplings lead to SGWBs with a large amplitude today, of the order of $h^2\Omega_{\rm GW}^{(0)} \simeq 5\cdot10^{-9}$. These backgrounds are however peaked at high frequencies $f_{\rm p} \sim 10^6-10^8$ Hz, which makes them undetectable by current/planned GW observatories. As the amount of GWs produced is in any case remarkable, we discuss the prospects for probing the SGWB indirectly by using constraints on the effective number of relativistic species in the universe $\Delta N_{\rm eff}$.}

\begin{document}
\maketitle
\flushbottom

\section{Introduction}

Significant evidence~\cite{Akrami:2018odb} supports the idea of inflation as a solution to the shortcomings of the hot Big Bang framework~\cite{Guth:1980zm, Linde:1981mu,Starobinsky:1980te}, and 
as a mechanism to create the primordial density perturbations~\cite{Mukhanov:1981xt,Guth:1982ec,Starobinsky:1982ee, Hawking:1982cz,Bardeen:1983qw} (see~\cite{Lyth:1998xn,Riotto:2002yw,Bassett:2005xm,Linde:2007fr,Baumann:2009ds} for reviews on inflation). Inflationary models must be compatible with cosmological observations~\cite{Martin:2013tda,Planck:2018jri}, including the most recent constraint on the B-mode polarization of the Cosmic Microwave Background (CMB), which sets an upper bound on the inflationary Hubble scale as $H_{\rm inf} \lesssim 4.7\times10^{13}$ GeV~\cite{BICEP:2021xfz}. This rules out many scenarios, and puts pressure on the parameter space of many others.   

Inflation must be followed by a period of \textit{reheating} during which the Universe ultimately has to reach a radiation dominated (RD) thermal state, at least before the onset of Big Bang Nucleosynthesis (BBN) at a temperature of $T_{\rm BBN} \simeq 10^{-3}\text{ GeV}$ \cite{Kawasaki:1999na,Kawasaki:2000en,Hannestad:2004px,Hasegawa:2019jsa}. The first stage of reheating may be driven by a period of \textit{preheating}, characterized by strong non-perturbative field excitation, typically leading to exponentially growing particle number densities. A paradigmatic example of this is {parametric resonance}, a phenomenon by which particle species coupled to the inflaton -- the {\it daughter} fields --, are created in energetic bursts due to the oscillations of the inflaton around the mininum of its potential. In the case of bosonic particles, the production of daughter species is resonant, and the energy transferred into them grows exponentially within few inflaton oscillations~\cite{Traschen:1990sw,Kofman:1994rk,Shtanov:1994ce,Kaiser:1995fb,Kofman:1997yn,Greene:1997fu,Kaiser:1997mp,Kaiser:1997hg}. Other mechanisms considered include particle creation due to trilinear or higher-order interactions between the inflaton and daughter field(s)~\cite{Dufaux:2006ee,Croon:2015naa,Antusch:2015vna,Enqvist:2016mqj}, geometric preheating of fields non-minimally coupled to gravity~\cite{Bassett:1997az,Tsujikawa:1999jh,Fu:2019qqe,Figueroa:2021iwm}, tachyonic preheating after Hybrid inflation of Higgs-like fields~\cite{Felder:2000hj,Felder:2001kt,GarciaBellido:2002aj,Copeland:2002ku} and gauge fields~\cite{Rajantie:2000nj,Copeland:2001qw,Smit:2002yg,GarciaBellido:2003wd,Tranberg:2003gi,Skullerud:2003ki,vanderMeulen:2005sp,DiazGil:2007dy,DiazGil:2008tf,Dufaux:2010cf,Tranberg:2017lrx}), gauge field excitation due to shift-symmetric interactions with an axion-like inflaton~\cite{Adshead:2015pva,Cuissa:2018oiw,Adshead:2019lbr,Cui:2021are}, etc. Multi-field (p)reheating scenarios have been also considered~\cite{Bezrukov:2008ut,Garcia-Bellido:2008ycs,Braden:2010wd,Giblin:2010sp,Figueroa:2022iho}, including the case with non-minimal gravitational couplings~\cite{DeCross:2015uza,DeCross:2016fdz,DeCross:2016cbs,Krajewski:2018moi,Iarygina:2018kee}. For reviews on (p)reheating we refer the reader to~\cite{Allahverdi:2010xz,Amin:2014eta,Lozanov:2019jxc,Allahverdi:2020bys}.

A common byproduct of early universe dynamics are gravitational waves (GWs). A (quasi-)scale invariant stochastic GW background (SGWB) is actually predicted in vanilla models due to quantum vacuum fluctuations~\cite{Grishchuk:1974ny,Starobinsky:1979ty, Rubakov:1982df,Fabbri:1983us,Saikawa:2018rcs,Kite:2021yoe}, whereas the dynamics of axion-like species during inflation generate blue-tilted signals~\cite{Anber:2006xt,Sorbo:2011rz,Pajer:2013fsa,Adshead:2013qp,Adshead:2013nka,Maleknejad:2016qjz,Dimastrogiovanni:2016fuu,Namba:2015gja,Ferreira:2015omg,Peloso:2016gqs,Domcke:2016bkh,Caldwell:2017chz,Guzzetti:2016mkm,Bartolo:2016ami}. SGWBs are also expected from post-inflationary phenomena, like particle production during preheating~\cite{Easther:2006gt,GarciaBellido:2007dg,GarciaBellido:2007af,Dufaux:2007pt,Dufaux:2008dn,Dufaux:2010cf,Bethke:2013aba,Bethke:2013vca,Figueroa:2017vfa,Adshead:2018doq,Adshead:2019lbr,Adshead:2019igv}, oscillon dynamics~\cite{Zhou:2013tsa,Antusch:2016con,Antusch:2017vga,Liu:2017hua,Amin:2018xfe}, kination-dynamics~\cite{Giovannini:1998bp,Giovannini:1999bh,Boyle:2007zx,Figueroa:2018twl,Figueroa:2019paj,Gouttenoire:2021jhk,Co:2021lkc}, and others. For reviews on SGWBs of cosmological origin see~\cite{Caprini:2018mtu,Maggiore:2018sht}. In this work, we focus on GW production during preheating. 

While many preheating studies have considered four-leg interactions $\phi^2\chi^2$ of the inflaton $\phi$ with another scalar field $\chi$, the
inflaton can never decay completely in such a case~\cite{Kofman:1997yn,Figueroa:2016wxr,Antusch:2020iyq,Antusch:2021aiw,Antusch:2022mqv}. To allow for a complete inflaton decay, interactions of the form $\phi \chi^n$ must be included. Trilinear interactions with $n = 2$ are simply the most natural choice, as interactions of this type are ubiquitous in particle physics. A paradigmatic example are Yukawa interactions, which represent a coupling between a scalar and two fermions. Three-leg decay via fermions was in fact the first channel of inflaton perturbative decay considered~\cite{Dolgov:1982th,Abbott:1982hn}, though it was realized later on that the inflaton also decays via non-perturbative parametric excitations in this case~\cite{Greene:1998nh,Greene:2000ew,Peloso:2000hy,Berges:2010zv}. Three-leg decay via interactions with bosons leads also to non-perturbative parametric excitations~\cite{Dufaux:2006ee}, and this channel is actually expected to dominate over Yukawa interactions, simply due to Pauli blocking of the Fermions. Furthermore, theories with spontaneous symmetry breaking or scalar theories charged under gauge symmetries, lead naturally to trilinear vertices between bosonic species, often involving gauge fields. Even if we restrict ourselves to scalar field interactions, trilinear
interactions are naturally expected in many contexts. A trilinear interaction $\phi\chi^2$ between an inflaton $\phi$ and a scalar daughter field $\chi$ represents therefore a natural coupling between an inflationary sector and a matter sector, and hence we focus on this type of interaction in this paper.

Considering in particular the interaction term
$\frac{1}{2}\sigma \phi \chi^2$, where $\sigma$ is a coupling of mass dimensions, we can understand the excitation mechanism underlying this interaction as follows. If following the end of inflation the inflaton oscillates coherently around the origin of its potential, the effective oscillatory frequency of a $\chi$-field mode function $\chi_{k}$, is given by $\omega_k^2 (t)=\frac{k^2}{a^2}+ m_\chi^2 + \sigma \, \phi \left( t \right)$, with $k$ the modulus of the comoving momentum, and $a(t)$ the scale factor. If the bare mass of $\chi$ satisfies $m_{\chi}^2 \ll \sigma\,|\phi|$, the trilinear coupling will induce a tachyonic mass for $\chi$ whenever $\phi < 0$, which happens during half of the time of each inflaton oscillation. Correspondingly, the modes within the infrared (IR)  window $k^2 < \sigma\, |\phi|\, a^2$ will be exponentially
amplified periodically, every half-period of the inflaton oscillations. This resonant mechanism, dubbed {\it tachyonic resonance} in Ref.~\cite{Dufaux:2006ee}, leads to a very efficient production of $\chi$ particles, so preheating concludes within few oscillations of $\phi$.  

If the daughter field $\chi$ is the Standard Model (SM) Higgs, the question of the stability of its vacuum becomes of upmost relevance. Even though the presence of a trilinear coupling does not affect the stability of the Higgs vacuum during inflation, it can influence it during preheating~\cite{Ema:2017ckf}. If the trilinear coupling $\sigma$ is large enough, the exponentially large occupation number of the Higgs created by tachyonic resonance can destabilize the vacuum. Lattice simulations have set upper bounds on the value of the trilinear coupling $\sigma$, and on the dimensionless coupling of a scale-free interaction $\frac{1}{2}\lambda_{h\phi}\phi^2 h^2$, based on demanding a stable evolution of the Higgs during preheating. According to Ref.~\cite{Enqvist:2016mqj}, Higgs vacuum stability is assured whenever the couplings respect the bounds $10^{-10}<\lambda_{h\phi}<10^{-8}$ and $\sigma < 2\times10^8$ GeV.     

In our present work we consider a daughter field $\chi$ with its self-interaction coupling strictly positive\footnote{In other words, $\chi$ could be identified with the SM Higgs only if the self-interaction coupling $\lambda$ of the Higgs remains positive, something that depends (in the absence of beyond the SM physics) very sensitively on the choice of $\alpha_{\rm s}$ and the top Yukawa coupling $y_{t}$~\cite{Degrassi:2012ry,Bezrukov:2012sa}.}. The excitation of the field modes $\chi_k$ within the aforementioned IR tachyonic window provides a non-trivial time-dependent spatial configuration for $\chi$, which in turn generates gravitational radiation. Here we study for the first time the GW production during preheating with a trilinear interaction $\frac{1}{2}\sigma \phi \chi^2$. 
We consider different inflaton potentials $V_{\rm inf}(\phi)$ with minimum at the origin, and study the GW production during preheating in each case as a function of $\sigma$. As the periodic tachyonic excitation of the IR modes $\chi_k$ resembles to some extent the excitation of gauge fields during preheating after axion-inflation~\cite{Adshead:2015pva,Cuissa:2018oiw}, and the latter is known to source GWs very efficiently~\cite{Adshead:2019igv,Adshead:2019lbr,Cui:2021are}, we expect as well a very efficient GW production due to the trilinear coupling. We investigate therefore whether we may obtain SGWBs such that their energy density could be constrained by upper bounds on the number of relativistic species in the universe~\cite{Planck:2018vyg,Pagano:2015hma,Abazajian:2019eic,COrE:2011bfs,EUCLID:2011zbd}, hence placing constraints on $\sigma$. Furthermore there is also the possibility that due to the nature of the trilinear interaction, the frequencies of the expected SGWB could be shifted towards smaller (more observable) frequencies than the SGWBs produced in standard $\phi^2\chi^2$ preheating~\cite{Figueroa:2017vfa}.

The paper is divided as follows. In Section~\ref{sec: mono}, we analyse the preheating dynamics and GW production for a monomial inflaton potential. In Section~\ref{sec:Poly}, we study the preheating dynamics and GW production for a {\it polynomial inflaton} potential. In Section~\ref{sec:Discussion}, we discuss our results and their potential implications. Some technical details are given in the appendices.\vspace*{0.1cm}

{\it Conventions --.} We consider a spatially-flat Friedmann-Lemaitre-Robertson-Walker (FLRW) background metric, $ds^2 = -dt^2 + a(t)^2\delta_{ij}dx^idx^j$, where $a(t)$ is the scale factor, and $t$ the cosmic time. We assume summation over repeated indices. On top of the background we consider tensor metric perturbations that satisfy $h_{ii} = \partial_i h_{ij} = 0$, hence identified with GWs. We denote the reduced Planck mass by $m_p = 2.435\cdot 10^{18}$ GeV.

\section{Monomial potential}\label{sec: mono}

We start our investigation by considering a quadratic potential for the inflaton $\phi$ around the origin. While a purely monomial potential $V_{\rm inf}(\phi) \propto \phi^2$ for inflation is strongly disfavored\footnote{Monomial potentials during inflation are not ruled out if an appropriate non-minimal coupling between the inflaton and the Ricci scalar is also present~\cite{Tsujikawa:2013ila}.} by B-mode CMB data~\cite{BICEP:2021xfz}, we can consider consistent inflaton potentials that flatten out towards large field values by developing a plateau, and then adopt a monomial shape around the origin only after inflation ends. These potentials are inspired by $\alpha$-attractor models~\cite{Kallosh:2013hoa} and can be written as $V_{\rm inf}\left(\phi\right)=\frac{1}{2}\Lambda^{4}\tanh^{2}\left({\phi/M}\right)$,
where $M$ and $\Lambda$ constants have dimensions of mass. In the limit $|\phi| \ll M$, the potential becomes $V_{\rm inf}\left(\phi\right) \simeq {1\over 2}m_{\phi}^2\phi^2$, with $m_{\phi}^2 \equiv \Lambda^2/M$. Compatibility with CMB observations at CMB scales allow us to choose $M = 5 m_p$, which then leads to $\Lambda = 0.00564m_p$, and from there to an inflaton mass $m_{\phi} \simeq 1.6\cdot 10^{13}$ GeV (see e.g.~Section II of Ref.~\cite{Antusch:2021aiw} for further details). Finding $m_{\phi}$ from fitting CMB anisotropy observables with an inflationary quadratic potential $V_{\rm inf}(\phi) = {1\over 2}m_{\phi}^2\phi^2$ at CMB scales, leads however to a very similar scale for the inflaton mass as in the $\alpha$-attractor case. So, in practice, we then simply consider a quadratic potential for the inflaton $\phi$, and supplement it with a trilinear interaction $\phi\chi^{2}$
with a daughter scalar field $\chi$, for which we allow a self-interaction $\propto\chi^{4}$,
\begin{equation}
V\left(\phi,\chi\right)=\frac{1}{2}\,m_{\phi}^{2}\,\phi^{2}+\frac{\sigma}{2}\,\phi\,\chi^{2}+\frac{\lambda}{4}\chi^{4},\label{pot mono}
\end{equation}
where $\sigma$ is a coupling with dimensions of mass, and $\lambda$ is the self-interaction coupling of the daughter field. We neglect the term $\frac{1}{2}\,m_{\chi}^{2}\,\chi^{2}$ in Eq.~(\ref{pot mono}) as we anticipate that the typical momenta excited $k/a \lesssim \sigma|\phi|$ will be very large, and hence we simply assume that $m_{\chi} \ll \sigma|\phi|$. It is however necessary to include the $\chi^{4}$ self-interaction to ensure that the potential is bounded from below and, consequently only stable dynamics are developed.
We then identify a {\it critical} coupling  
\begin{equation}
\lambda_c(\sigma,m_\phi) \equiv \frac{\sigma^{2}}{2\,m_{\phi}^{2}}\:,    \label{bounded}
\end{equation}
so that values $\lambda < \lambda_c$ are not allowed because the potential in Eq.~(\ref{pot mono}) would not be bounded from below. The critical value $\lambda_c$ defines in fact a flat direction $\phi=-\frac{\sigma\,\chi^{2}}{2\,m_{\phi}^{2}}$,
where $V\left(\phi,\chi\right)=0$. For $\lambda>\lambda_c$ the flat direction is lifted and the potential has a minimum at
$\phi=\chi=0$ where $V\left(\phi,\chi\right)=0$. The requirement
that the potential has to be bounded from below thus imposes an important
relation between the two couplings of the model, $\sigma$ and $\lambda$.
Throughout our following analysis, we adopt a value $\lambda = 2\lambda_c$ so that our numerical dynamics are guaranteed to avoid running-away solutions. 

\subsection{Preheating dynamics. Lattice simulations}
\label{sec: preheat mono and lattice}

The dynamics of preheating depend on the curvature of the potential
around its minimum, so our results hold for a class of inflationary
models with similar effective inflaton potential around the origin, as long as the leading term is quadratic. As mentioned before, the
monomial inflaton potential that we considered in Eq.~\eqref{pot mono}, $V_{\rm inf}\left(\phi\right)=\frac{1}{2}\,m_{\phi}^{2}\,\phi^{2}$,
is simply a very good approximation to the $\alpha$-attractor potential $V_{\rm inf}\left(\phi\right)=\frac{1}{2}\Lambda^{4}\tanh^{2}\left({\phi/M}\right)$ at field amplitudes
$\left|\phi\right|\ll M$. For simplicity in our analysis, we will fix the initial amplitude $\Phi_{\rm i}$ of the inflaton at the beginning of preheating (onset of inflaton oscillations), by determining the end of inflation in a quadratic inflaton potential $V_{\rm inf}(\phi)\propto\phi^2$. Following the end of inflation, the inflaton (in the form of a homogeneous condensate) starts oscillating around the
origin. During the oscillatory stage, the inflaton produces copiously particles of the $\chi$ field, transferring exponentially energy into the new quanta created. In our model, preheating occurs via a
trilinear interaction, which 
provides an effective time-dependent mass term for
$\chi$ with oscillatory sign in time. Whenever $\phi<0$, which occurs during half the time of each inflaton oscillation, the trilinear coupling
provides a negative squared mass for $\chi$. Hence, the corresponding
$\chi_k$ modes will be exponentially enhanced periodically, every time that $\phi<0$. This mechanism, known as tachyonic resonance, leads to a very efficient production of $\chi$ particles after few oscillations of $\phi$~\cite{Dufaux:2006ee,Enqvist:2016mqj}. 

In order to understand tachyonic resonance in more detail, we start by analyzing the equations of motion of the fields $\phi$
and $\chi$,
\begin{eqnarray}
\ddot{\phi}+3\,H\,\dot{\phi}+m_{\phi}^{2}\phi+\frac{\sigma}{2}\,\chi^{2}-\frac{{\nabla}^{2}\phi}{a^{2}}\, &=& 0\:, \label{eqn:eomm2p22}\\
\ddot{\chi}+3\,H\,\dot{\chi}+\lambda\,\chi^{3}+\sigma\,\chi\,\phi-\frac{{\nabla}^{2}\chi}{a^{2}} &=& 0\:,\label{eqn:eomm2p2}
\end{eqnarray}
where $a(t)$ is the scale factor and $H \equiv \dot a/ a$ the Hubble rate. It is convenient to perform a change of variables
\begin{eqnarray}
d\tilde{x}^{\mu}=m_{\phi}\,dx^{\mu}\,,~~~\tilde{\phi}=\frac{\phi}{\Phi_{i}}\,,~~~ \tilde{\chi}=\frac{\chi}{\Phi_{i}}\,,~~~\tilde{H}=\frac{H}{m_{\phi}}\:,
\end{eqnarray}
%
%
and then move into Fourier space, where we can naturally quantize the field $\chi$ as usual, writing
\begin{equation}\label{eqn:modeseqn}
    \tilde{\chi}({\bf x},t) = \int \frac{d^3 {\bf k}}{(2\pi)^3}[\hat{a}_k \tilde{\chi}_k(t) e^{i{\bf k\cdot x}} + \hat{a}^{\dagger}\tilde{\chi}_k^*(t)e^{-i{\bf k\cdot x}}]\:,
\end{equation}
with $\hat{a}_k^{\dagger}$, $\hat{a}_k$ standard creation/annihilation operators that satisfy $[\hat{a}_k,\hat{a}^{\dagger}_{k'}]=(2\pi)^3\delta({\bf k} - {\bf k}')$. The evolution equation of the mode functions, $\tilde{\chi}_k$, can be written as a linear equation using the Hartree approximation ${\tilde{\chi}^{3}\rightarrow3\,\tilde{\chi}\,\left\langle \tilde{\chi}^{2}\right\rangle}$ in Eq.~(\ref{eqn:eomm2p2}), so that
\begin{equation}
\tilde{\chi}_{k}''+3\,\tilde{H}\tilde{\chi}_{k}'+\left(3\,q_{\chi}\left\langle \tilde{\chi}^{2}\right\rangle +q_{3}\,\tilde{\phi}\left(t\right)+\frac{\kappa^{2}}{a^{2}}\right)\,\tilde{\chi}_{k}=0\,,~~~~~{\rm with}~~ \kappa={k\over m_{\phi}}
\label{Fourier space chi}\:,
\end{equation}
where prime $'$ denotes derivatives with respect to the new time variable, and the variance $\left\langle \tilde{\chi}^{2}\right\rangle$ can be related to the {\it power spectrum} of $\chi$ as follows
\begin{equation}
    \Varc = \int d \log k \:\Delta_{{\chi}}(k)~~,~~\Delta_{{\chi}}(k) \equiv \frac{k^3}{2\pi^2}\mathcal{P}_{{\chi}}~~,~~\langle{\chi}_{\bf k}{\chi}_{{\bf k}'}^*\rangle \equiv (2\pi)^3\mathcal{P}_{{\chi}}(k)\delta({\bf k} - {\bf k}')\:.
\end{equation}
We note that in Eq.~(\ref{Fourier space chi}) two dimensionless {\it resonance parameters} have naturally emerged,
\begin{equation}
q_{3}\equiv\frac{\sigma\,\Phi_{i}}{m_{\phi}^{2}}\quad\text{and}\quad q_{\chi}\equiv\frac{\lambda\,\Phi_{i}^{2}}{m_{\phi}^{2}}\:,\label{q3 and qchi def}
\end{equation}
with $\Phi_{i}$ the initial amplitude of the homogeneous field at the of inflation. These parameters, as we will see later, play a fundamental role in
our model, as they control the intensity of the daughter field excitation. Looking at the mode frequency in Eq.~\eqref{Fourier space chi}, we can write
\begin{equation}\label{mode equation}
\tilde{\omega}_{k}^{2} = m_{\tilde{\chi}}^{2}+(\kappa/a)^{2}\,,~~~~m_{\tilde{\chi}}^{2} = \left(q_{3}\,\tilde{\phi}\left(t\right)+3q_{\chi}\langle\tilde{\chi}^{2}\rangle\right).
\end{equation}
As the inflaton field oscillates in a quadratic potential as $\tilde{\phi}\left(t\right) \simeq \Phi(t)\sin\left(m_{\phi}\,t\right)$, with $\Phi(t)\propto {1\over t}\propto{1\over a^{3/2}}$~\cite{Turner:1983he}, we see that literally during every half oscillation the inflaton amplitude becomes negative. Hence, from Eq.~\eqref{mode equation}, we can see that every time that $\phi < 0$, $\tilde{\omega}_{k}^{2}$ becomes negative for sufficiently infrared momenta with modulus below the cut-off $k^{2}<-m_{\tilde{\chi}}^{2}$. This leads to an exponential growth of such infrared modes $\tilde\chi_k$, which translates at the same time into an exponential growth of the variance $\langle \tilde\chi^2 \rangle$ (or equivalently into an exponential growth of the power spectrum $\mathcal{P}_{{\chi}}(k)$ of the IR modes). Such exponential growth must actually happen in a step-like manner, as every time the inflaton amplitude is negative we expect the variance to grow, whereas every time the inflaton is positive we expect the variance to oscillate (simply due to the mode oscillations when $\tilde{\omega}_{k}^{2} > 0$ for $\phi> 0$). An overall exponential growth in steps is then expected.

In order to study the non-linear dynamics of the subsequent stages of the field excitation, we make use of lattice simulations. In this work, we use~\CL~\cite{Figueroa:2020rrl, Figueroa:2021yhd}. 
We have considered the set of equations Eq.~\eqref{eqn:eomm2p22} and Eq.~\eqref{eqn:eomm2p2} and solved numerically a discrete version of them. The initial conditions for the homogeneous part of the inflaton field during inflation is obtained by solving the Friedmann equation $H^2 \equiv \left({\dot a\over a}\right)^2 = \frac{1}{3 m_p^2}\left(\frac{1}{2}\dot{\phi}^2 + V(\phi)\right)$, together with Eq.~\eqref{eqn:eomm2p22} in the absence of the $\chi$ field (as well as in the absence of the $\nabla^2\phi$ term). We obtain the initial amplitude and velocity from the breaking of slow roll condition, $\epsilon_H \equiv -\frac{\dot{H}}{H^2} = 1$, what leads to
\begin{equation}\label{eqn:InitCondsM2Phi2}
    \Phi_i \equiv \phi(t_i) = 1.00738\:m_p \quad \text{and} \quad \dot{\Phi}_{i} \equiv \dot{\phi}\left(t_{i}\right) = -0.712295~~m_{\phi}\,m_p \: .
\end{equation}
The initial condition for the $\chi$ field are such that the homogeneous amplitude is set to zero, and standard quantum vacuum fluctuations are added on top, following the standard recipe for initial lattice power spectra, see e.g.~section~7 of Ref~\cite{Figueroa:2020rrl} for details.

\begin{figure*}[t] \begin{center}
    \includegraphics[width=0.45\textwidth,height=4.5cm]{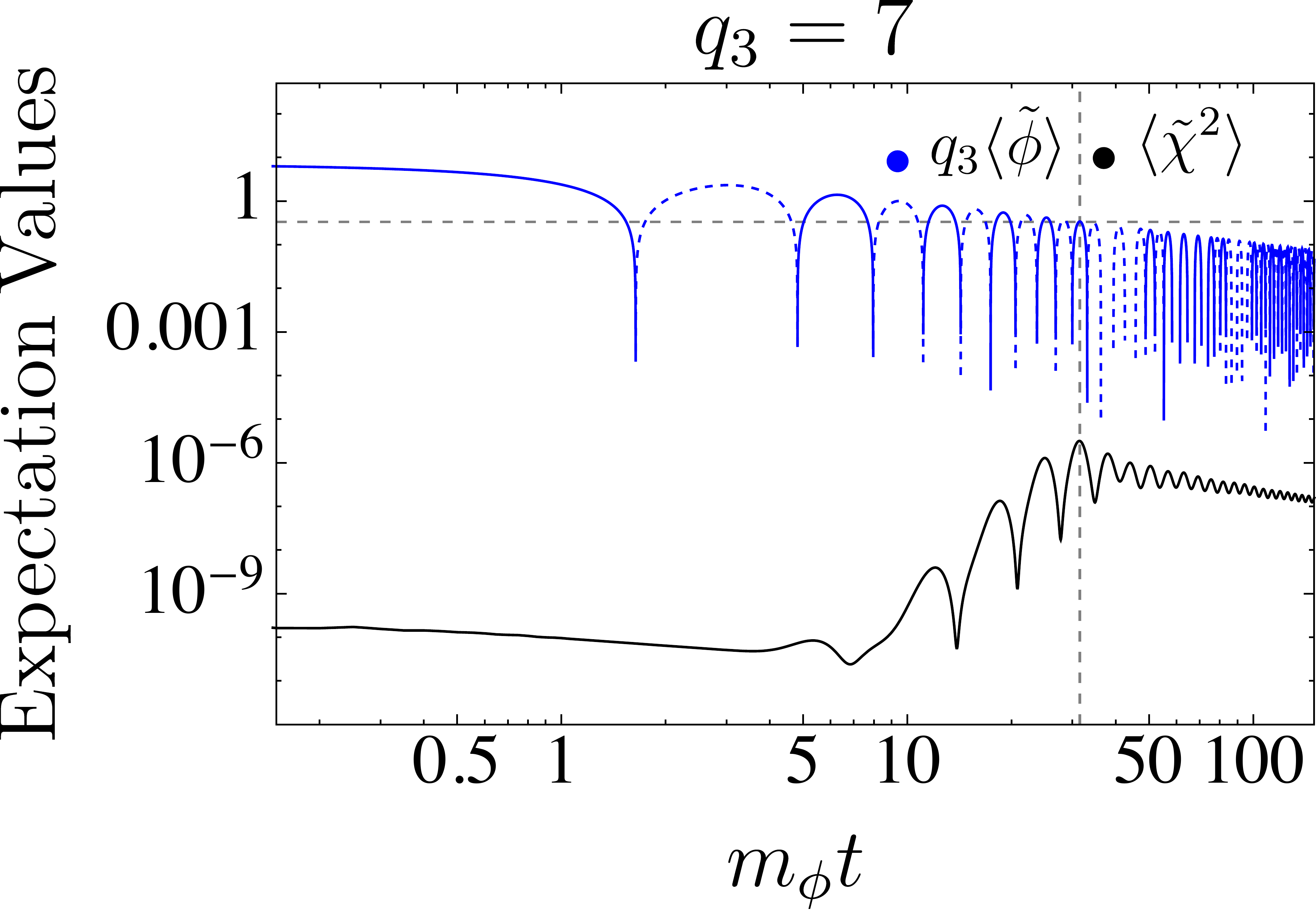}
    \includegraphics[width=0.45\textwidth,height=4.5cm]{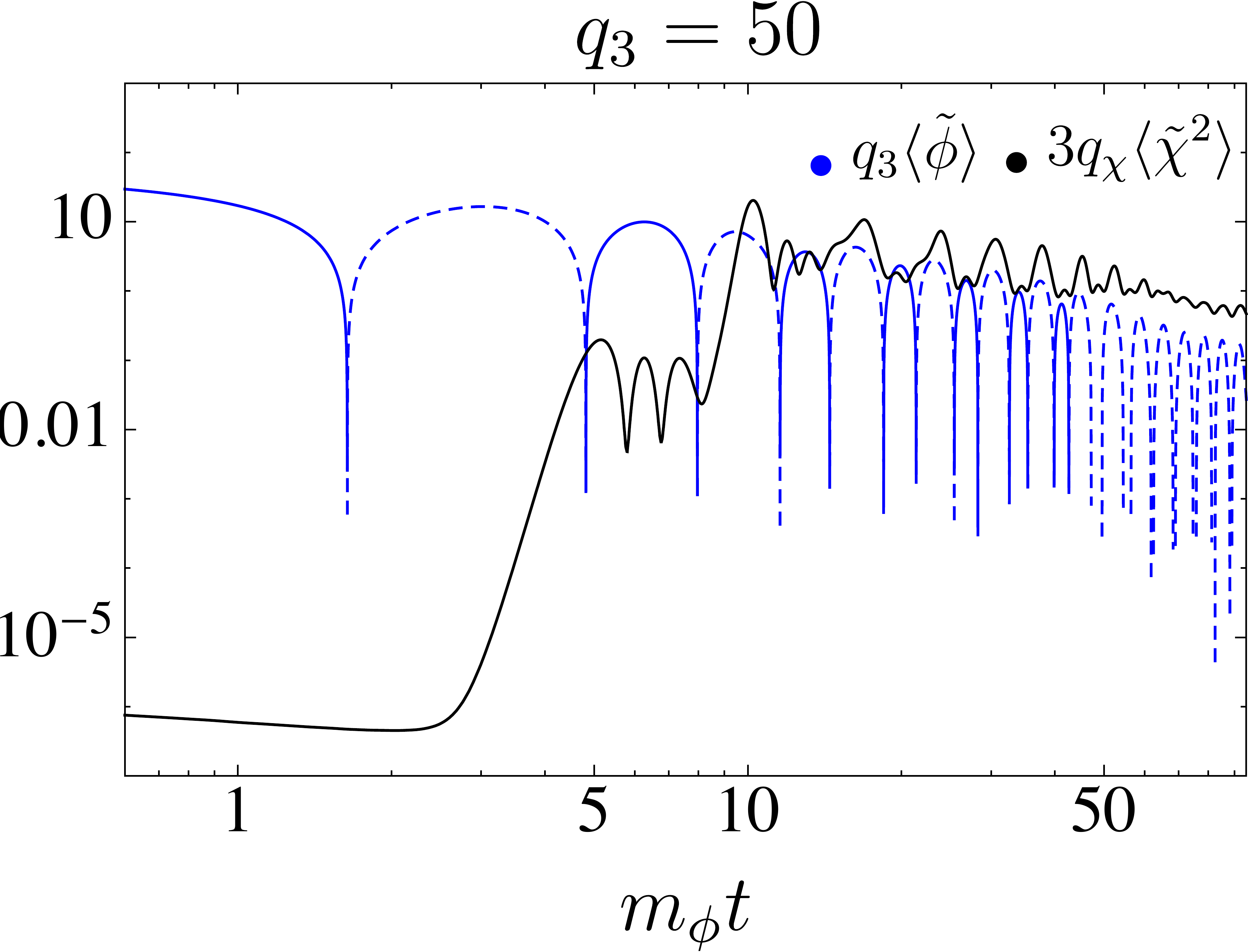} 
    \end{center}
    \vspace*{-0.5cm}
     \caption{{\it Left}: Time evolution of the inflaton's volume average $\langle\tilde{\phi}\rangle$ times the resonance parameter $q_3$ ($q_{3\:\text{eff}}=q_{3}\langle \tilde{\phi}\rangle$), and the variance of the daughter field $\Varc$ for $q_3=7$. The horizontal dashed line corresponds to $q_{3\:\text{eff}}=1/3$, whereas the vertical dashed line represents the time at which the maximum value of the variance is reached, $\Varc_{max}$. {\it Right:} Time evolution of $q_3 \langle \tilde{\phi}\rangle$ and the variance of the daughter field $\Varc$ times $3\,q_{\chi}$ for $q_3=50$. We see that once the term $3\,q_{\chi}\Varc$ overtakes $q_3 \langle \tilde{\phi}\rangle$, the growth of the variance stops. For both panels, the solid (dashed) blue lines correspond to the positive (negative) part of the inflaton mean value.} \label{fig:PhiVarchi} \vspace*{-0.3cm}
\end{figure*}

We have performed a series of simulations with $N=512$ lattice sites per dimension and a minimum infrared momentum $\kappa_{\rm IR}\in [0.2:0.75]$, where $\kappa_{\rm IR}=\frac{2\pi}{L}$ with $L$ the comoving length of the lattice, making sure we capture well all relevant scales of the problem. Simulations were run for different values of $q_3$ while considering the relation $q_\chi = q_3^2$ for $q_3 \in [1:10^4]$. Such relation between the resonance parameters simply implies that we are using $\lambda = 2\lambda_c$, in order to be sufficiently safe from run-away solutions in the numerical dynamics. Simulations were forced to end at time $m_{\phi} t_{\rm f} = 400$, as by this time in all cases (independently of the coupling $\sigma$), the system has reached a steady state with no further energy exchange between fields.

We now explain the dynamics we identify in our lattice simulations. An initial stage is characterized by the oscillations of the inflaton,  which remains as a homogeneous field for as long as the backreaction of the $\chi$ field can be neglected. In this regime, the solution to the equation of motion of the inflaton is ${\phi}(t) \simeq \sqrt{\frac{8}{3}}\frac{m_p}{m_{\phi}t}\cdot\sin(m_{\phi}t)$~\cite{Kofman:1997yn}. The effective frequency of oscillation of the modes $\chi_k$, as given in Eq.~\eqref{Fourier space chi}, is negative whenever the inflaton is negative, which induces an exponential growth of the modes below the cut-off
\begin{equation}\label{eqn:kband}
\kappa < \sqrt{q_{3} |\tilde{\phi}|}\, a\:.    
\end{equation}
Looking at the mode frequency squared $\tilde{\omega}_{k}^{2}$ in Eq.~(\ref{mode equation}),
we naturally identify an effective tachyonic resonance parameter as $q_{3,\text{eff}} = q_3 |\tilde{\phi}|$, with $q_3$ the initial resonance parameter defined in Eq.~(\ref{q3 and qchi def}). Due to the expansion of the universe, the inflaton envelope amplitude $\Phi(t)$ decays in time, and hence the effective resonance parameter decreases in time following the same decay as $q_{3,\text{eff}} \propto |\Phi| \propto {1 \over a^{3/2}}$. Once the value of $q_{\text{eff}}$ drops below $1/3$, the system switches to a narrow resonance regime~\cite{Dufaux:2006ee}, where the tachyonic excitation is no longer effective. Hence, depending on the value of the trilinear coupling, two regimes can be identified: $i)$ For 'small' values of $q_3$, $q_{3,\text{eff}}$ drops below $1/3$ before or soon enough after the first zero-crossing of the inflaton, and hence the growth of the variance of $\chi$ shuts down at a maximum value $\Varc_{max}$ determined by the moment when $q_{3,\text{eff}} = 1/3$, as shown in the {\it Left} panel of Fig.~\ref{fig:PhiVarchi}. $ii)$ For 'large' values of $q_3$, the variance $\Varc$ grows exponentially in steps (as many as negative semi-cycles of the inflaton) up to a new larger maximum value $\Varc_{max}$, which is reached before $q_{3,\text{eff}}$ drops below 1/3. We refer to this new maximum as the {\it critical} value of the variance, $\Varc_{max} = \Varc_{crit}$, see {\it Right} panel of Fig.~\ref{fig:PhiVarchi}. In this second case, just about when the variance reaches $\Varc_{crit}$, the backreaction of the daughter field into the inflaton cannot be ignored any longer. In particular, the growth of $\Varc$ towards $\Varc_{crit}$ displaces the position of the minimum of the inflaton potential, so the inflaton is 'locked' to be around a negative value after the variance of $\chi$ reaches $\Varc_{crit}$.

Numerically, we find the division between the two regimes to be at around $q_3 \simeq 10$, i.e.~the regime $i)$ of 'small' $q_3$ values corresponds to $q_3 < 10$, whereas the regime $ii)$ of 'large' $q_3$ values corresponds to $q_3 > 10$. Although the maximum value $\Varc_{max}$ corresponds in both regimes to the end of the tachyonic resonance, in the regime $i)$ such value is reached when the condition $q_{3,{\rm eff}} \leq 1/3$ is satisfied, whereas in the regime $ii)$ the maximum value corresponds to the moment when the $\chi$ self-interaction overtakes the trilinear interaction for the first time in $\chi$'s effective mass $m_{\tilde{\chi}} = q_3 \langle\tilde{\phi}\rangle+3q_{\chi}\Varc$. By cancelling out these two terms with each other (when $\tilde \phi < 0$), we can obtain an analytical estimation for the critical value $\Varc_{crit}$ in the regime $ii)$ as
\begin{equation}\label{eqn:Varcrit}
    \Varc_{crit} \simeq \dfrac{q_3 |\langle \tilde{\phi} \rangle|}{3 q_\chi }\,,~~~~(q_3 \gtrsim 10)\,.
\end{equation}
In Fig.~\ref{fig:PhiVarchi}, we show the time evolution of the mean value of the inflaton times the resonance parameter, $q_3\langle \tilde{\phi}\rangle$, and the variance of the daughter field $\Varc$, for $q_3=7$ (\textit{Left} panel) and $q_3=50$  (\textit{Right} panel), and for $q_\chi$ values that satisfy the relation $q_\chi = q_3^2$. Using the definitions of $q_{3}$ and $q_{\chi}$ in Eqs.~(\ref{q3 and qchi def}), we can easily see that the critical relation between $\sigma$ and $\lambda$ in Eq.~\eqref{bounded} translates into $q_{\chi}$=$\frac{q_{3}^{2}}{2}$. Thus, the relation $q_{\chi}=q_{3}^{2}$ used
represents simply a conservative choice to avoid a possible runaway
unstable solution of the dynamics. As expected, we observe that $\Varc$ grows every negative semi-oscillation, whereas it oscillates during the positive ones. The variance stops growing once it reaches its maximum value $\Varc_{max}$.

\begin{figure*}[t] \begin{center}
    \includegraphics[width=0.6\textwidth,height=5.5cm]{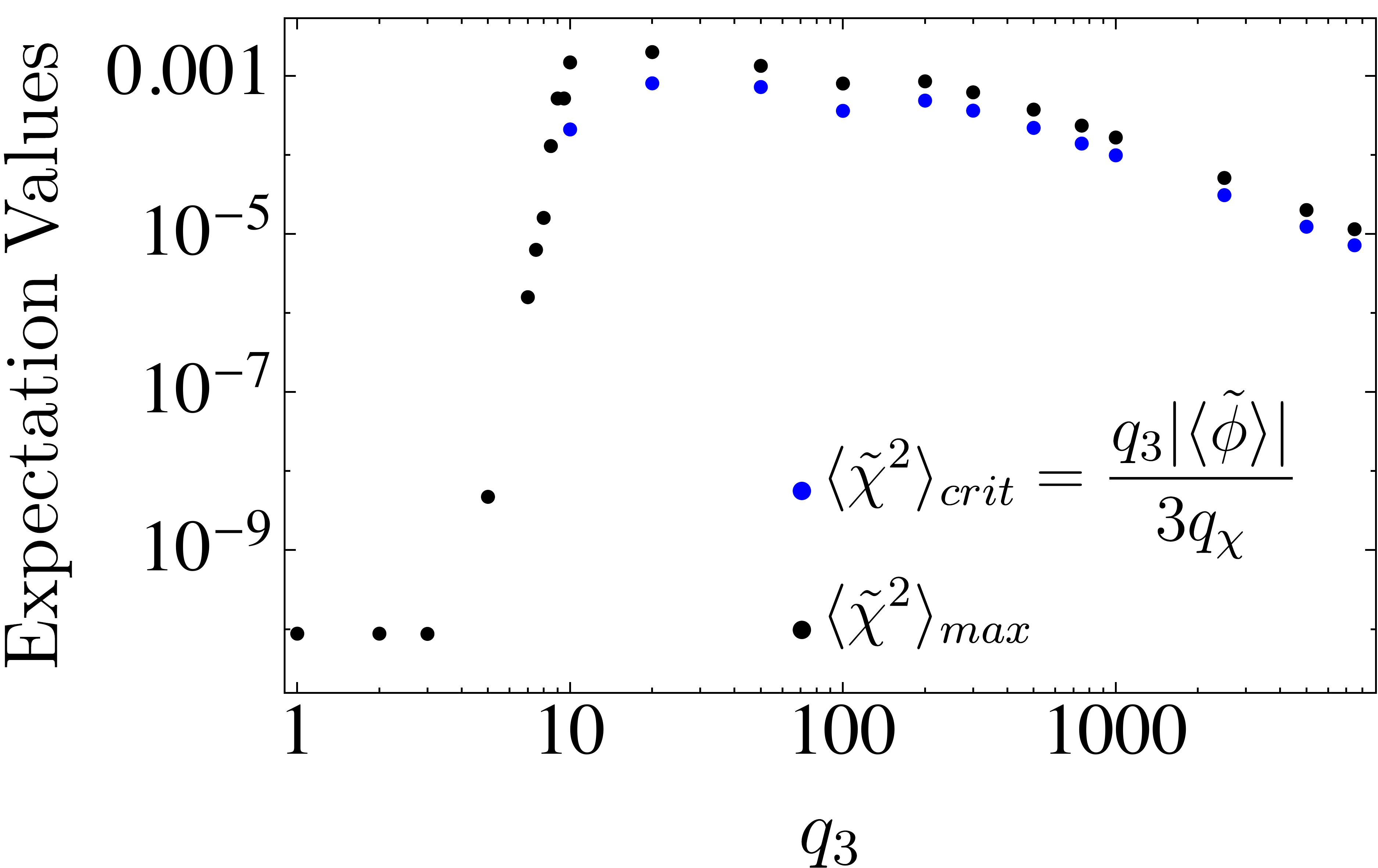} \hspace{0.5cm}
    \end{center}
    \vspace*{-0.5cm}
    \caption{Lattice results of $\Varc_{max}$ for different values of $q_3$ (black dots). It is clearly observed that for $q_3 \lesssim 10$ the amplitude of $\Varc_{max}$ is much smaller than for $q_3 \gtrsim 10$, as the tachyonic growth of $\Varc$ is simply stopped by the condition $q_{3,{\rm eff}} = 1/3$. We choose $\Varc_{max}$ as the value of $\Varc$ averaged over one oscillation when it reaches its maximum value. For comparison, we also show the corresponding theoretical prediction for $q_3 \gtrsim 10$ in Eq.~\eqref{eqn:Varcrit} (blue dots).}\label{fig:PhiVarchicrit}\vspace*{-0.3cm}
\end{figure*}

In Fig.~\ref{fig:PhiVarchicrit}, we plot the maximum values $\Varc_{max}$ extracted from lattice simulations, as a function of the resonance parameter $q_3$, and compare them against our theoretical estimation for the critical value $\Varc_{crit}$ in the regime $ii)$ with $q_3 \gtrsim 10$, c.f.~Eq.~\eqref{eqn:Varcrit}. 
Recalling the relation $q_{\chi}=q_{3}^{2}$ we used in our simulations, we can see that the critical variance for $q_3 \gtrsim 10$ decreases inversely proportional to $q_3$, as $\Varc_ \simeq \langle \tilde{\phi} \rangle / (3q_3)$, hence inversely proportional to the coupling constant $\sigma$. 
Although our analytical prediction Eq.~\eqref{eqn:Varcrit} only captures roughly the real amplitude $\Varc_{max}$\footnote{We estimate the maximum variance $\Varc_{max}$ as the average over one period of oscillation of the variance $\Varc$, once it has reached its maximum value.}, it captures exactly the dependence on $q_3$.

%

\subsection{Gravitational wave production}\label{sec:GWsM2Phi2}

\begin{figure*}[t] \begin{center}
    \includegraphics[width=0.75\textwidth,height=5.5cm]{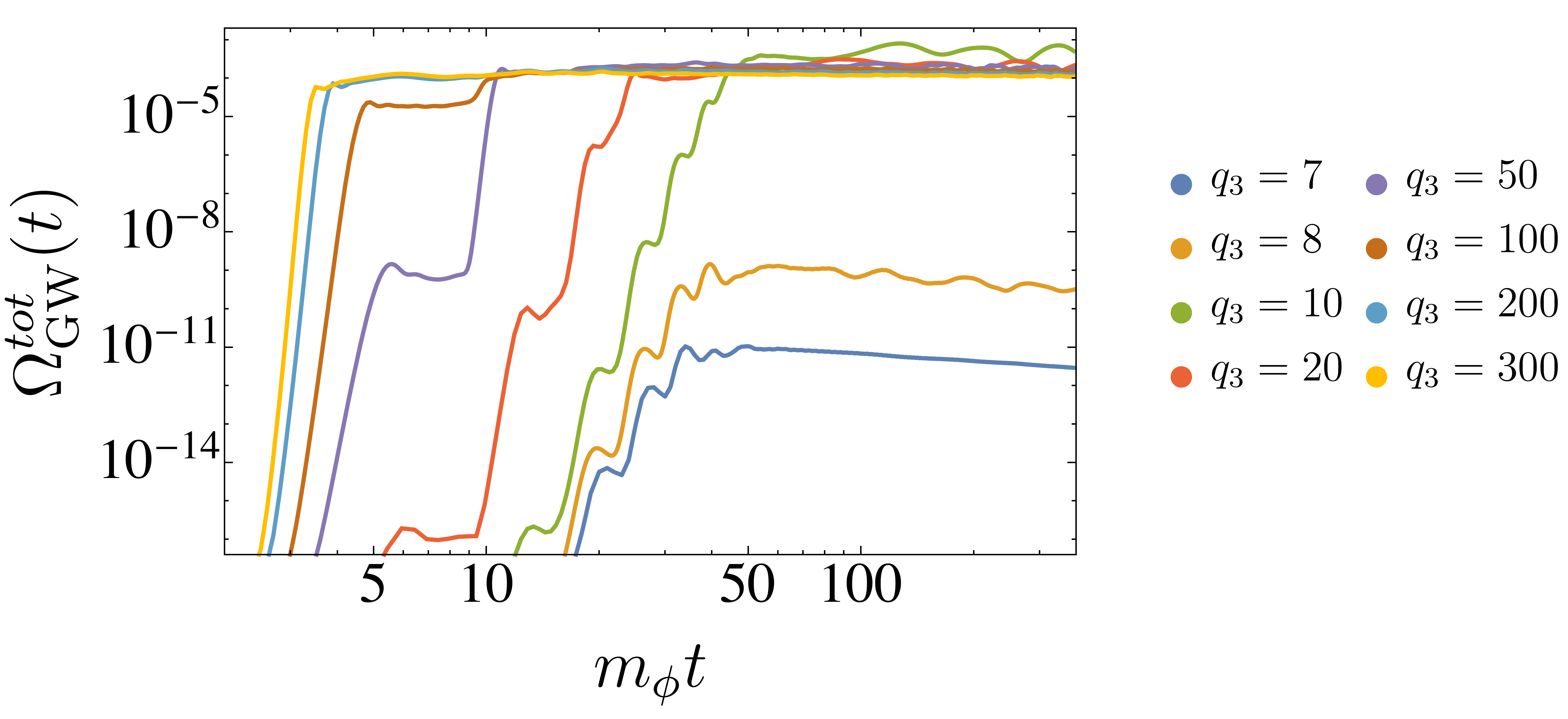} \hspace{0.5cm}
    \vspace*{-0.5cm}
    \end{center}
    \vspace*{-0.2cm}
    \caption{Time evolution of the GW total energy density for different values of $q_3$.
    } \label{fig:rhoGW} \vspace*{-0.3cm}
\end{figure*}

In this section, we study GW production during preheating using the recently released module of \CL~dedicated to the generation of GWs by scalar fields~\cite{CLGWs}. The dynamics of GWs are governed by the linearized equation of motion of the tensor perturbations $h_{ij}$,
\begin{equation}\label{eqn:GWsEoM}
    \ddot{h}_{ij} + 3 H \dot{h}_{ij} - \dfrac{\nabla^2}{a^2}h_{ij} = \dfrac{2}{m_p^2 a^2}\Pi_{ij}^{TT}\:,
\end{equation}
which satisfy $h_{ii}=\partial_i h_{ij}=0$. Here $\Pi_{ij}^{TT}$ is the transverse-traceless part of the effective anisotropic stress tensor $\Pi_{ij} = \{\partial_{i} \phi \partial_j  \phi  + \partial_{i} \chi \partial_j \chi \}$. In momentum space, the TT operation is defined as $\Pi^{TT}_{ij}({\bf k},t) = \Lambda_{ijkl}({\bf k}) \Pi_{ij}({\bf k},t)$, with the projector $\Lambda$ defined as
\begin{eqnarray}\label{eqn:ProjectorTT}
    \Lambda_{ij,lm}({\hat{\bf k}}) \equiv P_{il}({\hat{\bf k}})  P_{jm}({\hat{\bf k}}) - \dfrac12 P_{ij}({\hat{\bf k}}) P_{lm}({\hat{\bf k}}) \: ,~~~~ P_{ij} = \delta_{ij} - {{\hat{\bf k}}}_i {{\hat{\bf k}}}_j\:,\quad  {{\hat{\bf k}}}_i = {\bf{k}}_i/k \:. ~~~~~~~~~
\end{eqnarray}
The energy density power spectrum of GWs normalized by the critical energy density of the universe is
\begin{equation}\label{eqn:OmegaGW}
  \Omega_{\rm GW}(k,t) = \frac{1}{\rho_c}\dfrac{d \rho_{\rm GW}}{d\log k}({\bf k}, t) = \dfrac{k^3}{(4\pi)^3GV} \int  \dfrac{d \Omega_k}{4\pi} \dot{h}_{ij}({\bf k},t) \dot{h}^*_{ij}({\bf k},t) \: ,
\end{equation}
where $d\Omega_k$ represents a solid angle element in {\bf k}-space. Lattice simulations follow the procedure described in Ref~\cite{Garcia-Bellido:2007fiu}, by solving a discrete version of Eq.~\eqref{eqn:GWsEoM} (see appendix~\ref{app:GWs} for further details). In order to compute the GW energy density power spectrum we make use of a discrete version of Eq.~\eqref{eqn:OmegaGW}. To obtain the total energy density of GWs, we integrate Eq.~\eqref{eqn:OmegaGW} over
momentum $k$:
\begin{equation}
\Omega_{\rm GW}^{\rm tot}\left(t\right)=\int{d\log k}\,\frac{1}{\rho_{c}}\dfrac{d\rho_{{\rm GW}}}{d\log k}({\bf k},t).
\end{equation}
\begin{figure*}[t] \begin{center}
     \includegraphics[width=0.45\textwidth,height=5cm]{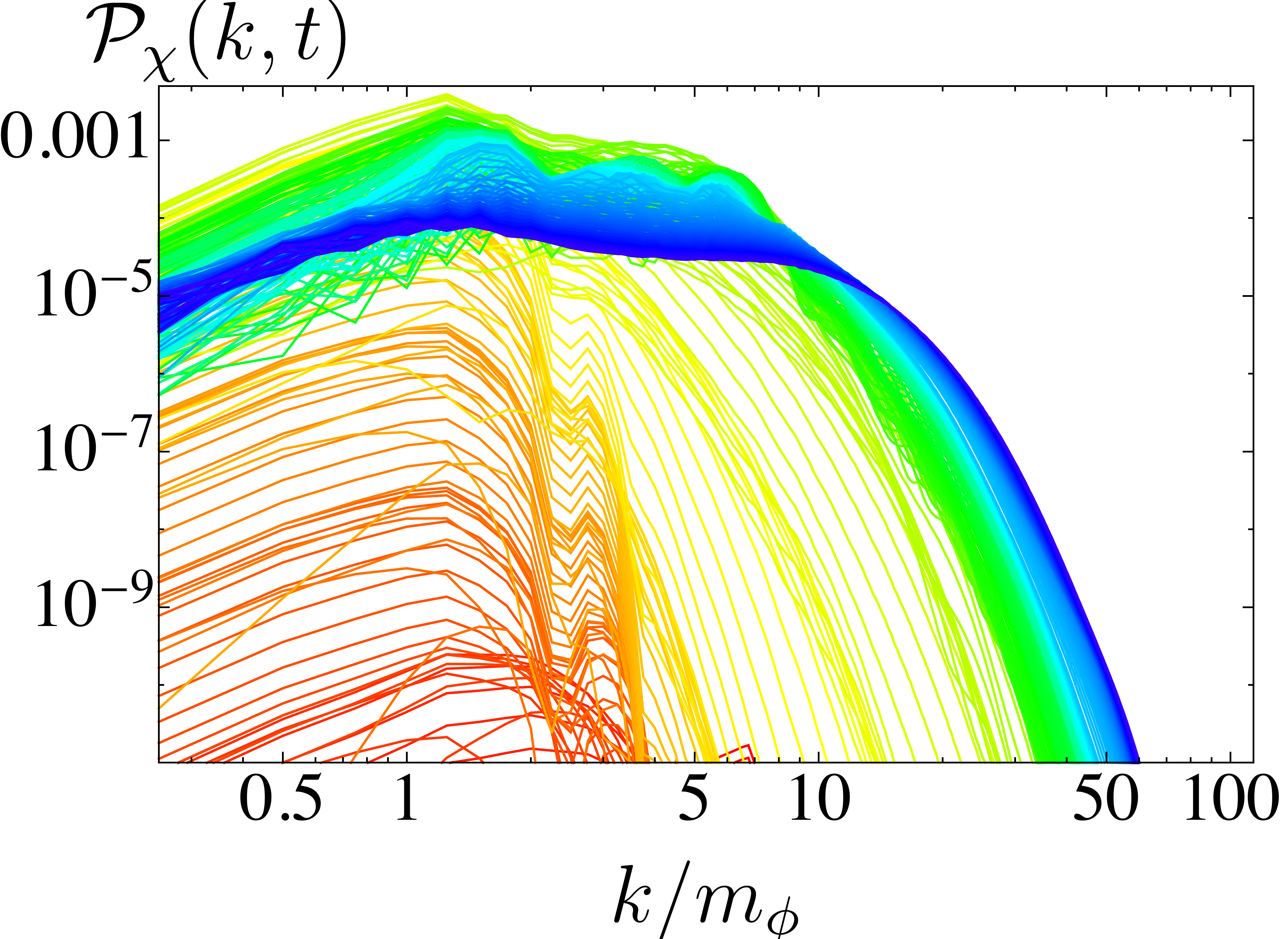} 
     \includegraphics[width=0.45\textwidth,height=5cm]{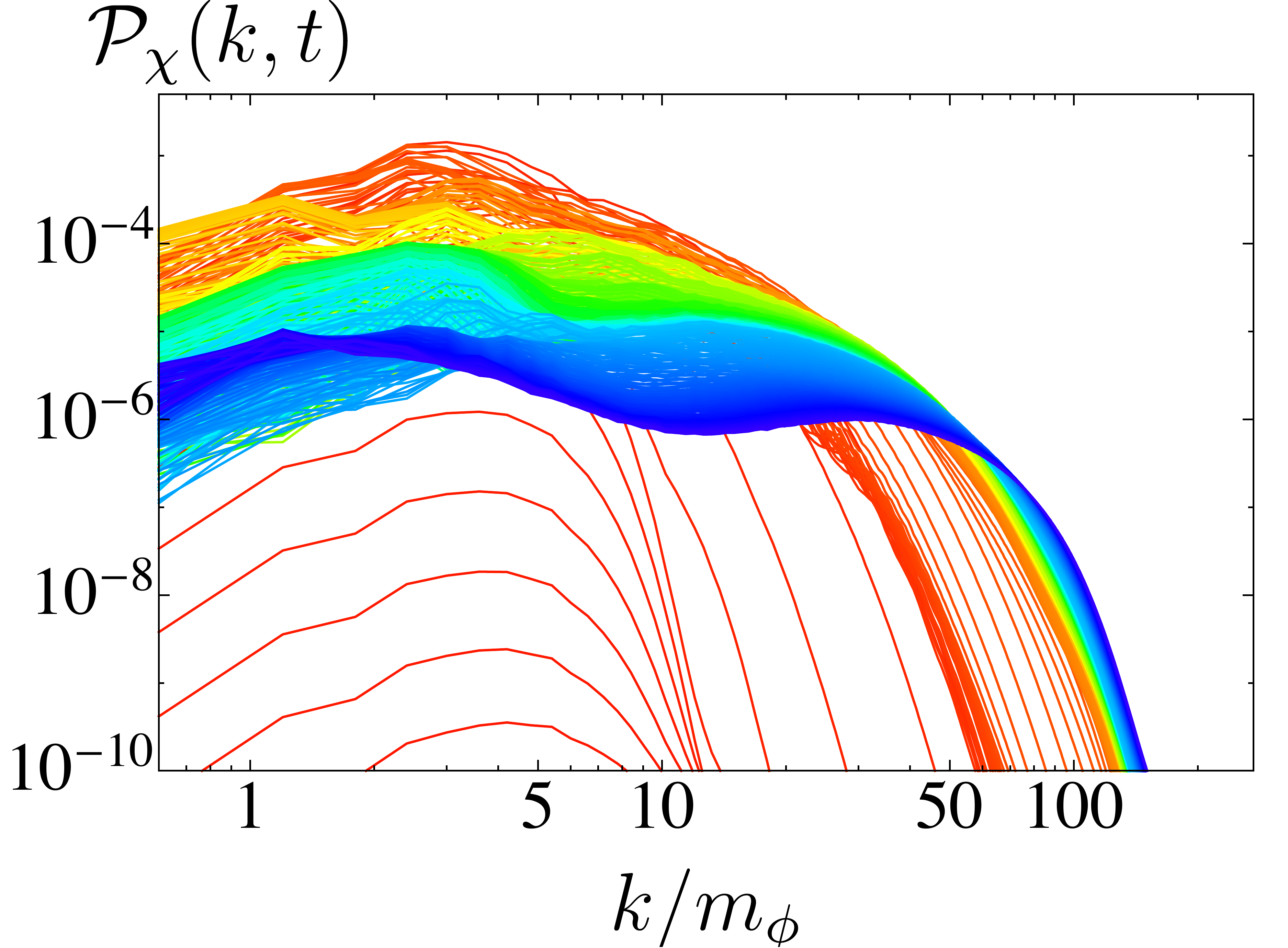} 
    \includegraphics[width=0.45\textwidth,height=5cm]{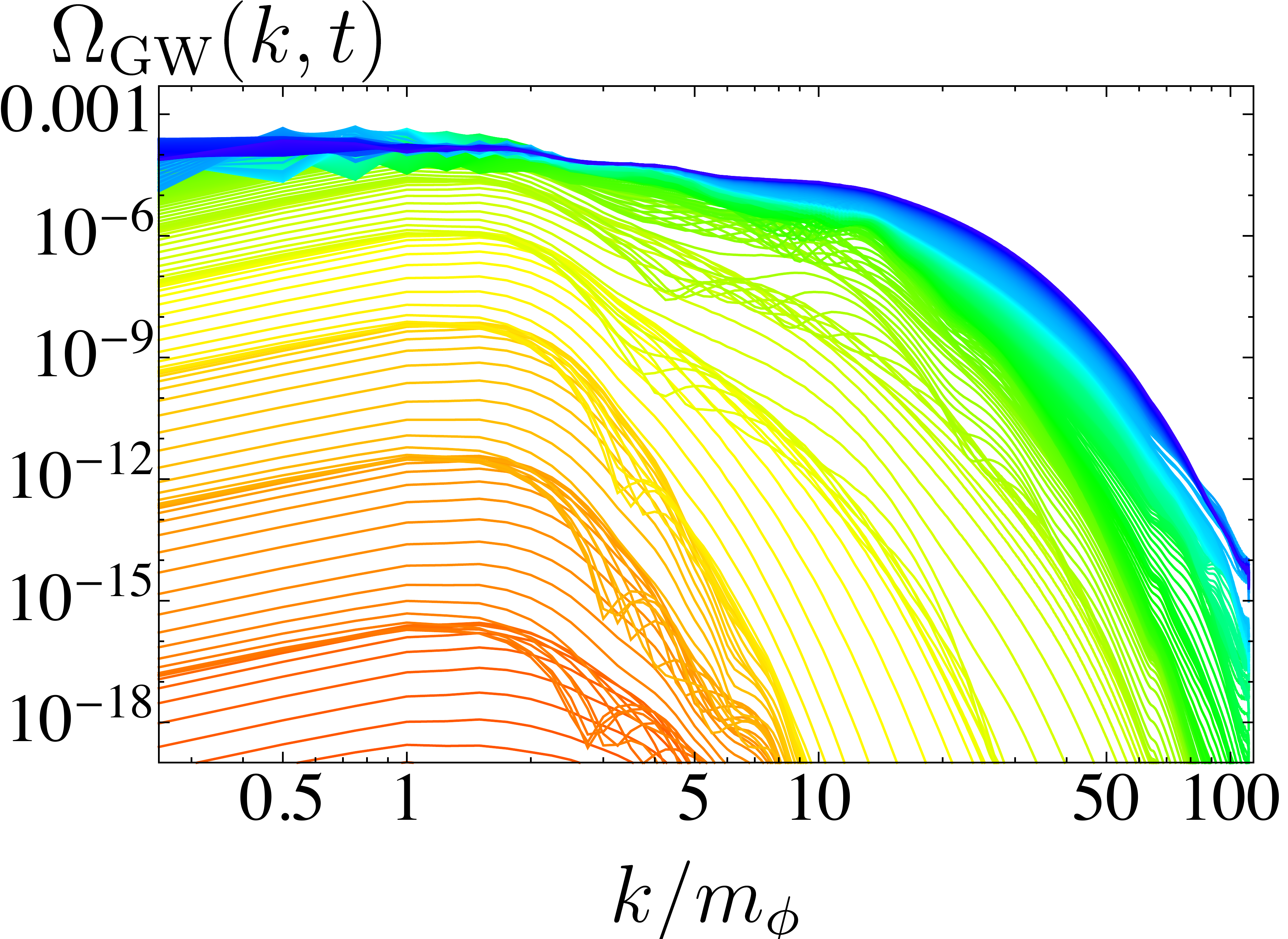} 
     \includegraphics[width=0.45\textwidth,height=5cm]{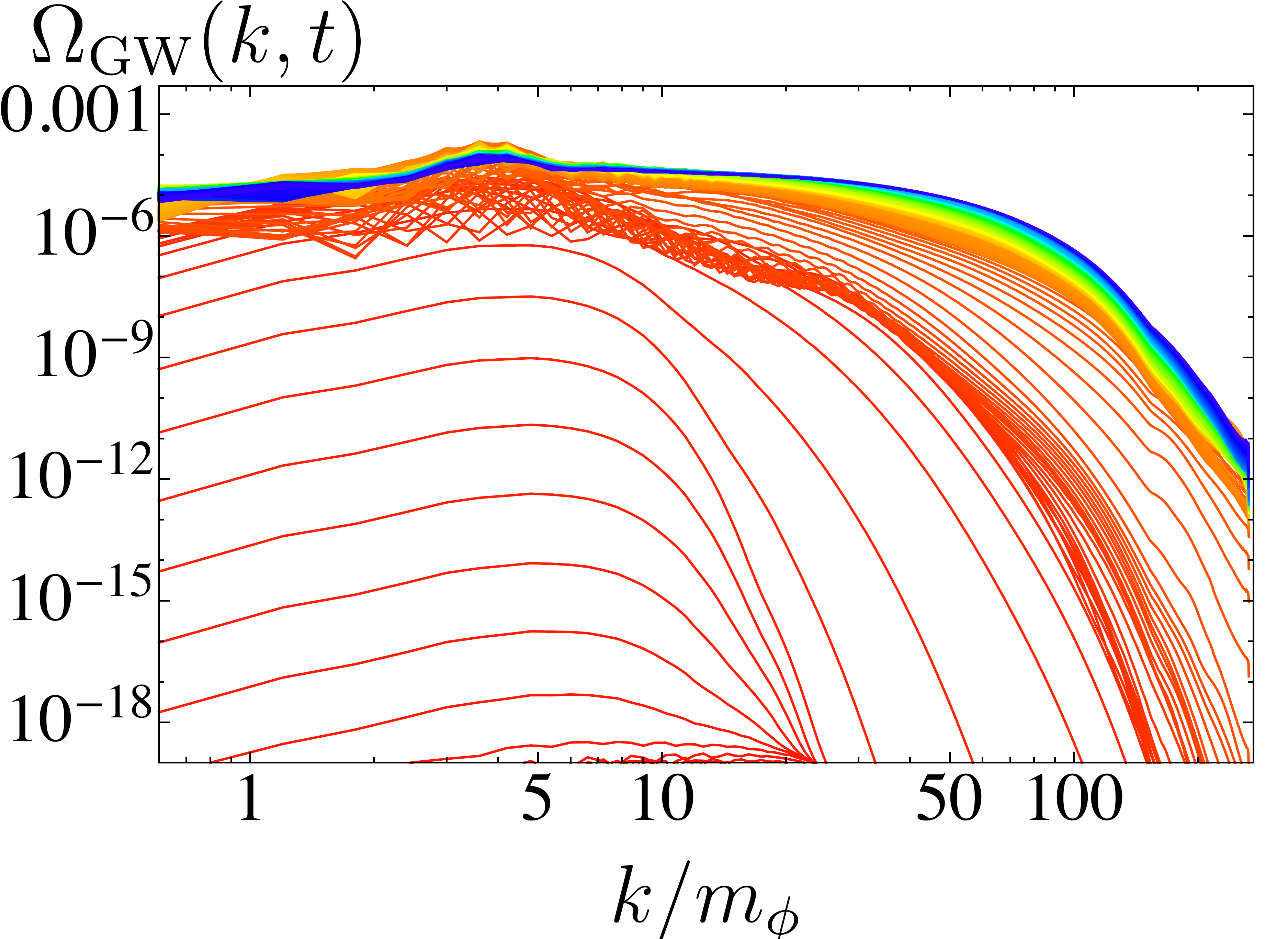} 
    \end{center}
    \vspace*{-0.5cm}
    \caption{{\it Top} panels: Matter power spectrum of the daughter field $\chi$, $\mathcal{P}_{\chi}(k)$, as a function of the momentum $k/m_{\phi}$ for $q_3=10$ ({\it Left}) and for $q_3=100$ ({\it Right}). {\it Bottom} panels: GW energy density power spectrum $\Omega_{\rm GW}(k)$, for $q_3=10$ ({\it Left}) and for $q_3=100$ ({\it Right}). In all panels, colors run from red (early times) to blue (late times). Each spectra is measured every $\Delta(m_\phi t)=0.2$.} \label{fig:ChiGWspectrum} \vspace*{-0.3cm}
\end{figure*}
In Fig.~\ref{fig:ChiGWspectrum}, we show in the {\it Top} panels the matter power spectrum  of the daughter field, $\mathcal{P}_{\tilde{\chi}}$, whereas in the {\it Bottom} panels we show the GW energy density power spectrum, $\Omega_{\rm GW}$, as a function of $\kappa = k/m_{\phi}$. Spectra are measured every $\Delta(m_{\phi} t) = 0.2$ up to a final time $m_{\phi} t_{\rm f} = 400$, assuring that the system has reached a steady state and the transfer of energy between $\phi$ and $\chi$ fields is negligible. The gravitational waves are produced in pulses coinciding with the periods of tachyonic instability of $\chi$, with step-like jumps visible in both the matter spectra $\mathcal{P}_{\tilde{\chi}}(k,t)$ and in $\Omega_{\rm GW}(k,t)$. The production of GWs ceases when the variance of the daughter field stops growing due to the end of tachyonic resonance (from that moment onward the $\chi_k$ modes are no longer excited, so the spatial gradients of $\tilde{\chi}$ do not evolve any further). In Fig.~\ref{fig:rhoGW} we show the evolution of the total GW energy density in time. We observe that it grows exponentially until it saturates to a certain value. The saturation occurs of course when $\Varc$ stops growing and reaches its maximum value $\Varc_{\max}$. For instance, we see that GW production for $q_3=50$ stops around $m_{\phi} t \sim 15$ and comparing this with the  {\it Right} panel of Fig.~\ref{fig:PhiVarchi}, we can observe that $\Varc$ stopped increasing around the same time. Once the GW spectrum saturates, the SGWB redshifts freely. Since the maximum value of $\Varc$ depends on $q_3$, the same dependence is inherited in the maximum value of the GW energy density, see Fig.~\ref{fig:rhoGW}. Another feature worth noticing is that for $q_{3}\gtrsim 10$ the final GW energy density ratio $\Omega^{tot}_{\rm GW}(t)$ reaches a maximum (approximately) fixed amplitude, which can be as high as $\Omega^{tot}_{\rm GW}(t_{\rm f}) \simeq 5\cdot 10^{-4}$ for 
$q_{3} \simeq 10$, while it decreases smoothly for larger values of $q_{3}$ [see Eq.~(\ref{eqn:fOmegaGWbis}) for a parametrization on the $q_3$ dependence of the peak amplitude of the SGWB].

\begin{figure*}[t] \begin{center}
    \includegraphics[width=0.55\textwidth,height=5.5cm]{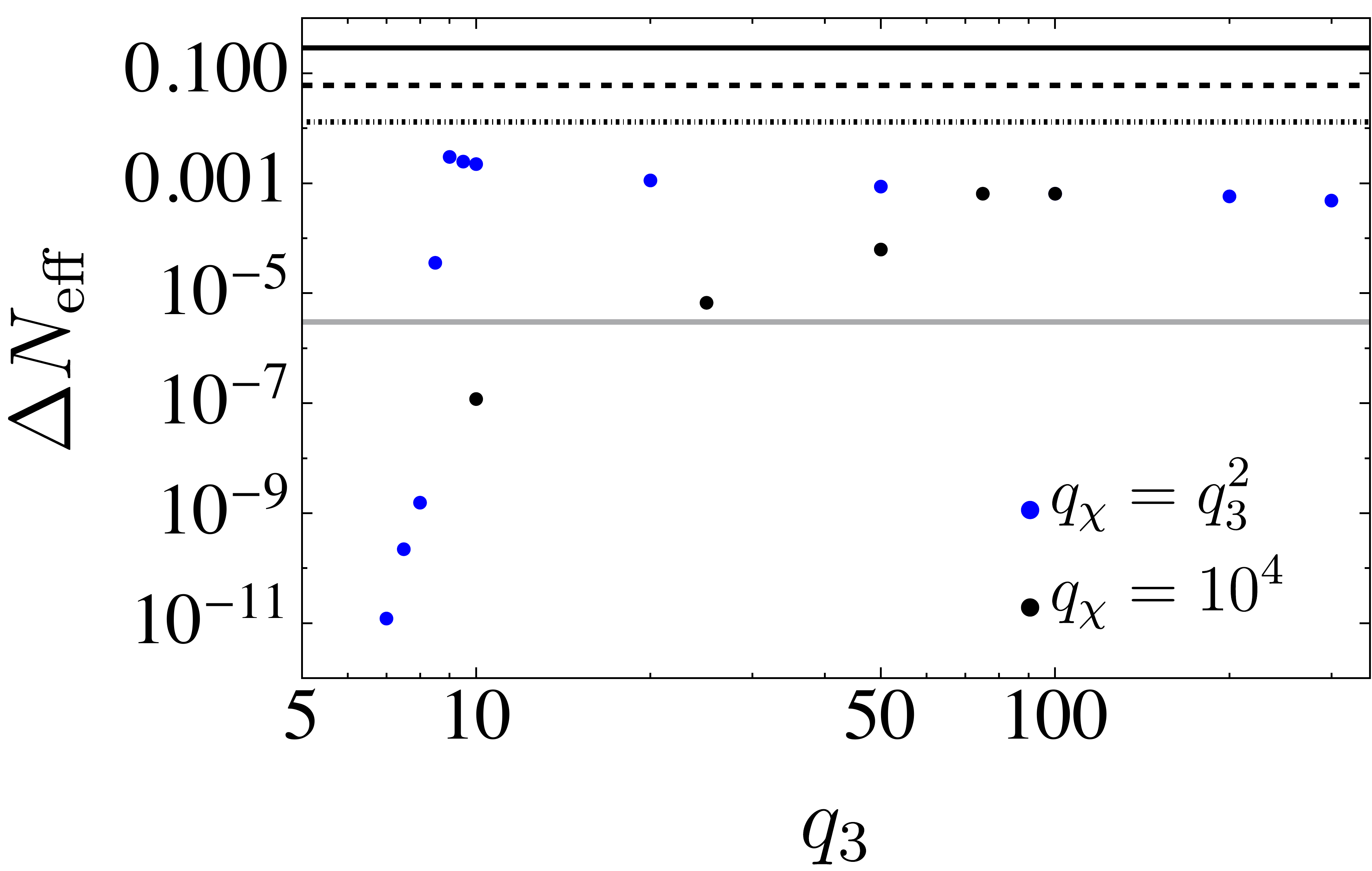} \hspace{0.5cm}
    \end{center}
    \vspace*{-0.2cm}
    \caption{Number of relativistic degrees of freedom beyond Standard Model for different values of $q_3$, for the relation $q_{\chi}=q_3^2$ and fixed $q_{\chi} = 10^4$. The solid top line corresponds to the current Planck bound $\Delta N_{\rm eff}\lesssim 0.29$, whereas the dashed line below represents the future CMB-S4 $\Delta N_{\rm eff} \lesssim 0.06$ upper bound at $2\sigma$. The dash-dotted line corresponds to the upper bound imposed by next generation satellite missions at $2\sigma$, $\Delta N_{\rm eff}\lesssim0.013$. The gray bottom solid line corresponds to a hypothetical cosmic variance limited (CVL) CMB polarization experiment upper bound  $\Delta N_{\rm eff}\lesssim 3\cdot10^{-6}$.
    } \label{fig:Neff} \vspace*{-0.3cm}
\end{figure*}

The SGWB produced during preheating is redshifted freely once its amplitude saturates. Today's amplitude and peak frequency of the SGWB depend on the expansion history of the universe, so in order to estimate them, we need to redshift the background from the final time of our simulations $t_{\rm f}$, till the present time $t_0$. This requires to know the expansion history from $t_{\rm f}$ till the moment $t_{\rm RD}$ when radiation domination (RD) is fully established. If we parametrize such intermediate era with an effective equation of state $\bar{w}=p/\rho$, we can write the redshifted frequency and amplitude today of the SGWB as (see appendix~\ref{app:Redshift} for further details)
\begin{eqnarray}
    f_{\rm GW} &\simeq& 4\times10^{10}\:\epsilon_{\rm f}^{1/4}\:\frac{k}{a_{\rm f} H_{\rm f}}\left(\frac{H_{\rm f}}{m_p}\right)^{1/2} \quad \text{Hz}\:,\label{eqn:redshiftfreq}\\
    h_0^2\Omega_{\rm GW}^{(0)} &\simeq& 1.6 \times 10^{-5} \: \epsilon_{\rm f} \: \Omega_{\rm GW}^{\rm (f)} \: ,\label{eqn:redshiftamp} 
\end{eqnarray}
where $\epsilon_{\rm f}$ is defined as
\begin{equation}
    \epsilon_{\rm f} \equiv \left(\frac{a_{\rm f}}{a_{\rm RD}}\right)^{1-3\bar{w}}\,,
\end{equation}
with $a_{\rm f} \equiv a(t_{\rm f})$ and $a_{\rm RD} \equiv a(t_{\rm RD})$. In order to estimate $h_0^2\Omega_{\rm GW}^{(0)}$ and $f_{\rm GW}$ we have characterized the position $\kappa_{\rm p}$ and the amplitude $\Omega_{\rm GW}^{\rm (f)}$ of the peak in the GW spectra at $t_{\rm f}$. While the position of the peak is characterized by the initial instability tachyonic band given in Eq.~\eqref{eqn:kband}, the final amplitude of the peak $\Omega_{\rm GW}^{\rm (f)}$ decreases smoothly with the resonance parameter. We have also measured the values of $a_{\rm f}$ and $H_{\rm f}$ in our simulations. Very importantly, we have observed that for $q_{3}\gtrsim 10$ the dominant contribution to the total energy of the system comes from the daughter field, which behaves as radiation. Therefore, the system at $t_{\rm f}$ is already quite close to RD, with an equation of state $\bar\omega$ slightly smaller but very close to $\bar\omega \simeq 1/3$. This implies that we can estimate $\epsilon_{\rm f}$ to be $\lesssim 1$. Putting everything together and taking $\epsilon_{\rm f} = 1$ for simplicity, we find the following parametrization for the SGWB peak today (valid only for $q_3\gtrsim 10$)
\begin{eqnarray}\label{eqn:fOmegaGW}
    f_{\rm GW}^{(p,0)}(q_3) &\simeq& (1.0 \pm 0.1)\times10^8 \: \left(\frac{q_3}{10}\right)^{0.52 \pm 0.04} \: \text{Hz}\: ,\\
    \label{eqn:fOmegaGWbis}
    h_0^2\Omega^{(p,0)}_{\rm GW}(q_3) &\simeq& (2.67 \pm 0.5)\times10^{-9} \: \left(\frac{q_3}{10}\right)^{-0.43 \pm 0.07}\,.
\end{eqnarray}
We observe that the produced SGWB has a rather large amplitude but is peaked at high frequencies. We also note that the exponent of our frequency characterization coincides with that in Eq.~\eqref{eqn:kband}. While a high-frequency detection program at $\sim$ MHz frequencies and above, has been recently put forward~\cite{Aggarwal:2020olq}, this is yet in its infancy, and the reality of the situation is that neither current nor planned GW detection experiments are so far expected to be able to detect these high frequency SGWBs. Direct detection is therefore challenging, as it is usually the case in $\phi^2\chi^2$ preheating scenarios also with monomial inflaton potentials. As the GW production is in any case very significant in our scenario, we may still look for another manner to probe these backgrounds, namely, using indirect constraining means.
In particular, as GWs produced during preheating have shorter wavelengths than the Hubble scale at the time of production, they contribute to the energy density of the universe as radiation~\cite{Caprini:2018mtu}. Hence, it might be possible to set bounds on the amount of GW energy density that can be produced, by using cosmological constraints on the total radiation density species allowed in the universe. This is typically parameterized by a deviation of the effective number of relativistic degrees of freedom beyond the Standard Model $\Delta N_{\rm eff} = N_{\rm eff} - N_{\rm eff,SM}$, with $N_{\rm eff,SM}=3.0440$~\cite{Bennett:2020zkv}. Assuming that only our SGWB contributes to the extra radiation energy density at the time of CMB decoupling, a bound on the extra radiation degrees of freedom $\Delta N_{\rm eff}$ sets a constraint on $h_0^2\Omega_{{\rm GW}}^{(0)}$ as~\cite{Caprini:2018mtu}
\begin{equation}\label{eqn:Neff}
    \frac{h_0^2\Omega_{{\rm GW}}^{(0)}}{h_0^2\Omega_{{\gamma}}^{(0)}}=\frac{7}{8}\left(\frac{4}{11}\right)^{4/3}\Delta N_{\rm eff}\:,
\end{equation}
where the present energy density of photons is $h_0^2\Omega_{{\gamma}}^{(0)} = 2.47 \times 10^{-5}$. The Planck observations set a limit $|\Delta N_{\rm eff}| \lesssim 0.29$ at 95\% C.L.~\cite{Planck:2018vyg,Pagano:2015hma}, whereas the next-generation CMB-S4 experiments will probe $\Delta N_{\rm eff} \lesssim 0.06$ at $2\sigma$~\cite{Abazajian:2019eic}. Additionally, the next generation of satellite missions like COrE~\cite{COrE:2011bfs} and Euclid~\cite{EUCLID:2011zbd} will impose bounds at $2\sigma$ on $\Delta N_{\rm eff}\lesssim 0.013$. A hypothetical cosmic variance limited (CVL) CMB polarization experiment could presumably get down to $\Delta N_{\rm eff}\lesssim 3\cdot10^{-6}$~\cite{Ben-Dayan:2019gll}, though this does not seem experimentally realistic. In Fig.~\ref{fig:Neff}, we show $\Delta N_{\rm eff}$ for different values of $q_3$ considering two cases: $q_{\chi}=q_3^2$ (blue dots) and $q_{\chi}=10^4$ (black dots). We found that the coupling that produces the largest amount of GWs is $q_3 = 9$, yielding $\Delta N_{\rm eff} \sim 0.003$, which is above the bound expected in a hypothetical CVL experiment, but still below the sensitivity of actually planned realistic experiments. 
We conclude that neither direct or indirect detection methods can probe the SGWBs predicted in the scenario studied in this section, as either the peak frequency of the GW signals is too high, or the amount of GW energy density produced is simply not enough to be constrained by $\Delta N_{\rm eff}$. In order to improve this situation, two circumstances could be sought for: 1) decreasing the energy scale of inflation, so that we can shift to smaller frequencies the SGWB peak, and/or 2) increasing the amount of GWs produced relative to the field energy budget in the universe. In the next section, we introduce a polynomial potential for the inflaton which, {\it a priori}, seems to allow to satisfy both conditions. We therefore explore next to what extent GW production during preheating after polynomial inflation can improve the detectability of the SGWB due to a trilinear interaction.


\section{Polynomial potential}
\label{sec:Poly}

In this section, we consider the most general renormalizable single-field model of inflation, where the inflaton potential is a polynomial of degree four \cite{Drees:2021wgd}
\begin{eqnarray}
V_{\rm inf}\left(\phi\right) &=& {1\over2}m_\phi^2\phi^2 + \sigma_\phi\phi^3 + \lambda_\phi\phi^4\nonumber\\ &=&\lambda_\phi\,\left(2\,\phi_{0}^{2}\,\phi^{2}-\frac{8}{3}\,\left(1-\beta\right)\,\phi_{0}\phi^{3}+\phi^{4}\right)\:,\label{eq:inflatonpoly}
\end{eqnarray}
where in the second line we have parametrized the mass and trilinear coupling of the inflaton as $m_\phi \equiv 2\sqrt{\lambda_\phi}\phi_0$ and $\sigma_\phi \equiv -{8\over3}\lambda_\phi(1-\beta)\phi_0$. The linear term in $\phi$ is absent in the potential of Eq.~\eqref{eq:inflatonpoly} since it can be absorbed by shifting the field, whereas the constant is set to the value of the cosmological constant today, which
is negligible when compared to the energy scale of inflation \cite{Drees:2021wgd}.
All three terms in Eq.~\eqref{eq:inflatonpoly} are used to match the observational data, and the existence of an (near) inflection-point at $\phi_{0}$ guarantees the flatness of the polynomial potential of quartic order. In this setup, $\lambda_\phi$ is the inflaton self-coupling, and $\beta$ is a parameter that controls the flatness configuration in the vicinity of $\phi_{0}$. Notice that, if $\beta<0$, there is a false vacuum for $\phi>\phi_{0}$, and the inflaton can get stuck in this second minimum \cite{Bernal:2021qrl}. Hence, in this work, we focus on the case where $0<\beta\ll1$, and on small inflaton amplitudes, so that the inflaton field value is sub-Planckian at the time when the curvature perturbations observed at the CMB are generated, i.e.,~we will assume $\phi_0\leq m_p$ \cite{Drees:2021wgd,Bernal:2021qrl}. Interestingly, the inflection-point, $\phi_{0}$, constitutes the only free parameter in this inflationary model, since all the couplings in the potential Eq.~\eqref{eq:inflatonpoly} can be expressed at its expense (for more details of this, see Refs. \cite{Drees:2021wgd,Bernal:2021qrl}). In particular, by considering the CMB measurements from Planck data \cite{Planck:2018jri} on
the scalar power spectrum, $P_{\zeta}=\left(2.1\pm0.1\right)\times10^{-9}$ and the spectral index $n_{s}=0.9649\pm0.0042$, assuming these were generated $N_{e}=65$ e-folds before the end of inflation, the quantities $\lambda_\phi$, $\beta$ and $m_\phi^2$, can be fixed as
\begin{eqnarray}
\lambda_\phi &\simeq& 6.61\times10^{-16}\,\left(\frac{\phi_{0}}{\mathrm{m_p}}\right)^{2}\:,\label{d}\\
\beta &\simeq& 9.73\times10^{-7}\,\left(\frac{\phi_{0}}{\mathrm{m_p}}\right)^{4}\:,\label{beta}\\
m_{\phi} &=& 2\,\sqrt{\lambda_{\phi}}\,\phi_{0}\simeq5\times10^{-8}\,\left(\frac{\phi_{0}}{m_{p}}\right)^{2}\,m_{p}\:.\label{eqn:mphipoly}
\end{eqnarray}
%
%
As a lower bound $\phi_0 \geq  3\times10^{-5}m_p$ emerges naturally by demanding the stability of the
inflationary scenario against radiative corrections \cite{Drees:2021wgd,Bernal:2021qrl}, the inflection point $\phi_{0}$ ranges in practice within the interval $3\times10^{-5}m_p < \phi_{0} \leq m_p$. The inflaton mass lies therefore in our modeling within the range $10^{2}\,\mathrm{GeV}<m_{\phi}\leq 10^{11}$ GeV. 

The full potential of the system inflaton-daughter field is then again
\begin{equation}
V\left(\phi,\chi\right)=V_{\rm inf}\left(\phi\right)+\frac{\sigma}{2}\,\phi\,\chi^{2}+\frac{\lambda}{4}\,\chi^{4}\:,\label{eq:pot2}
\end{equation}
and in order to ensure that this potential is bounded from below,
we require $\lambda_{\phi},\lambda>0$. Identifying a critical value for the self-coupling of $\chi$ as
\begin{equation}
\lambda_{c}\left(\sigma,\lambda_{\phi},\phi_{0}\right)\equiv\frac{\sigma^{2}}{8\,\lambda_{\phi}\,\phi_{0}^{2}}\,.\label{lambda pot2}
\end{equation}
We observe that for $0<\lambda < \lambda_c$, the potential~\eqref{eq:pot2} 
has two minima at non-zero $\chi$
and $\phi$ where the potential is negative, $V\left(\phi,\chi\right)<0$.
For $\lambda \geq \lambda_c$, there is
only one minimum at $\phi=\chi=0$ with $V\left(\phi,\chi\right)=0$.
In the following analysis we use the limiting case $\lambda = \lambda_c$, as we will see, it corresponds to the case for which the production of GWs is maximized.


\subsection{Preheating dynamics. Lattice simulations}
\label{sec:Preheating and Lattice Poly}

Similarly to the case of the monomial potential, we can perform a change of
variables such that we are able to write the equations of motion with dimensionless quantities. We choose conveniently
\begin{eqnarray}
d\tilde{x}^{\mu}=\sqrt{\lambda_{\phi}}\,\phi_{0}\,dx^{\mu}\,,~~~\tilde{\phi}=\frac{\phi}{\phi_0}\,,~~~ \tilde{\chi}=\frac{\chi}{\phi_0}\,,~~~\tilde{H}=\frac{H}{\sqrt{\lambda_{\phi}}\phi_{0}}\:,
\end{eqnarray}
such that the equations of motion for $\phi$ and $\chi$ become
\begin{eqnarray}
\tilde{\phi}''+3\,\tilde{H}\,\tilde{\phi}'+4\,\tilde{\phi}-8\left(1-\beta\right)\,\tilde{\phi}^{2}+4\,\tilde{\phi}^{3}+\frac{q_{3}}{2}\,\tilde{\chi}^{2}-\frac{\tilde{\nabla}^{2}\tilde{\phi}}{a^{2}} &=& 0\:, \label{eqn:polyphieom} \\
\tilde{\chi}''+3\,\tilde{H}\,\tilde{\chi}'+q_{3}\,\tilde{\chi}\,\tilde{\phi}+3\,q_{\chi}\,\left\langle \tilde{\chi}^{2}\right\rangle \,\tilde{\chi}-\frac{\tilde{\nabla}^{2}\tilde{\chi}}{a^{2}} &=& 0\:,\label{eq:polychieom}
\end{eqnarray}
where $'$ denotes derivatives with respect to the new time variable, and we have used again the Hartree approximation ${\tilde{\chi}^{3}\rightarrow3\,\tilde{\chi}\,\left\langle \tilde{\chi}^{2}\right\rangle}$.
The resonance parameters emerging now in the problem are defined as
\begin{equation}
q_{3}\equiv\frac{\sigma}{\lambda_{\phi}\,\phi_{0}}\quad\text{and}\quad q_{\chi}\equiv\frac{\lambda}{\lambda_{\phi}}\:.\label{eqn:respoly}
\end{equation}
The relation between, equivalent to $\lambda \geq \lambda_c$ so that  that there is only one minimum at  $\tilde{\phi}=\tilde{\chi}=0$ where $V\left(\tilde{\phi},\tilde{\chi}\right)=0$,
is given by
\begin{equation}
q_{\chi}>\frac{q_{3}^{2}}{8}\:.\label{qchip q3p}
\end{equation}
Introducing the field decomposition from Eq.~\eqref{eqn:modeseqn} into Eq.~\eqref{eq:polychieom}, we find that the mode functions of $\tilde{\chi}$ obey
\begin{eqnarray}
\tilde{\chi}_{k}''+3\,\tilde{H}\tilde{\chi}_{k}'+\left(3\,q_{\chi}\,\left\langle \tilde{\chi}^{2}\right\rangle +q_{3}\,\tilde{\phi}\left(t\right)+\frac{\kappa^{2}}{a^{2}}\right)\,\tilde{\chi}_{k}=0\label{poly Fourier space chi}\:,~~~{\rm where}~~\kappa \equiv \frac{k}{\sqrt{\lambda_{\phi}}\phi_0}
\end{eqnarray}
The $\chi$ field has therefore an effective time-dependent mass $m_{\tilde{\chi}}^{2}\approx\left(q_{3}\,\tilde{\phi}\left(t\right)+3q_{\chi}\langle\tilde{\chi}^{2}\rangle\right)$, which oscillates between negative and positive values due to the oscillations of the inflaton. Whenever $m_{\tilde{\chi}}^{2} < 0$ and $\kappa^{2}/a^2<-m_{\tilde{\chi}}^{2}$, the mode frequency $\tilde{\omega}_{k}^{2} \equiv \kappa^2/a^2 + m_{\tilde{\chi}}^{2}$ will become negative leading to an exponential growth of the corresponding $\tilde{\chi}_k$ modes. The power spectrum of the excited modes, $\mathcal{P}_{\tilde\chi}(k,t)$, and consequently the variance of the $\tilde\chi$ field, $\left\langle \tilde{\chi}^{2}\right\rangle$, will then grow exponentially in a step-like manner, every time the inflaton amplitude turns negative.

In order to follow in detail the dynamics of tachyonic resonance in this scenario, similarly as in section~\ref{sec: preheat mono and lattice}, we have considered Eqs.~\eqref{eqn:polyphieom} and~\eqref{eq:polychieom} and solved numerically a discrete version of these. We obtain the initial amplitude and velocity of the inflaton from the breaking of the slow-roll condition when $\epsilon_H = 1$, for different values of $\phi_0$: 
%
\begin{table}[htb!]
\renewcommand{\arraystretch}{1.15}
\addtolength{\tabcolsep}{5pt} 
\begin{centering}
\begin{tabular}{| c | c | c |}
\hline
$\phi_{0}$ & $\tilde{\phi}(t_{i})$ & ${\tilde{\phi}}'(t_{i})$\\
\hline 
{}&{}&{}\\[-3.25ex]\hline\\[-3.5ex] 
$m_p$ & 0.368 & -0.395 \\
\hline 
$10^{-1}\,m_p$ & 0.473 & -0.464\\
\hline 
$10^{-2}\,m_p$ & 0.485 & -0.470\\
\hline 
$10^{-3}\,m_p$ & 0.486 & -0.471\\
\hline 
\end{tabular}
\par\end{centering}
\caption{The initial amplitude and velocity of the inflaton for different values
of $\phi_{0}$.}
\end{table}
%
%
%

We have performed a series of lattice simulations with $N=512$ sites per dimension and a minimum infrared momentum $\kappa_{\rm IR}\in[0.5:6]$, making sure we capture well all relevant scales. Simulations were run for different values of $q_{3}$ considering the relation $q_{\chi}=q_{3}^2/8$ for $q_{3} \in [25:5\times10^4]$, finishing at $\sqrt{\lambda_{\phi}}\phi_0 t_{\rm f} = 400$, which guarantees that the system has reached a steady state. 

\begin{figure*}[t] \begin{center}
    \includegraphics[width=0.45\textwidth,height=4.75cm]{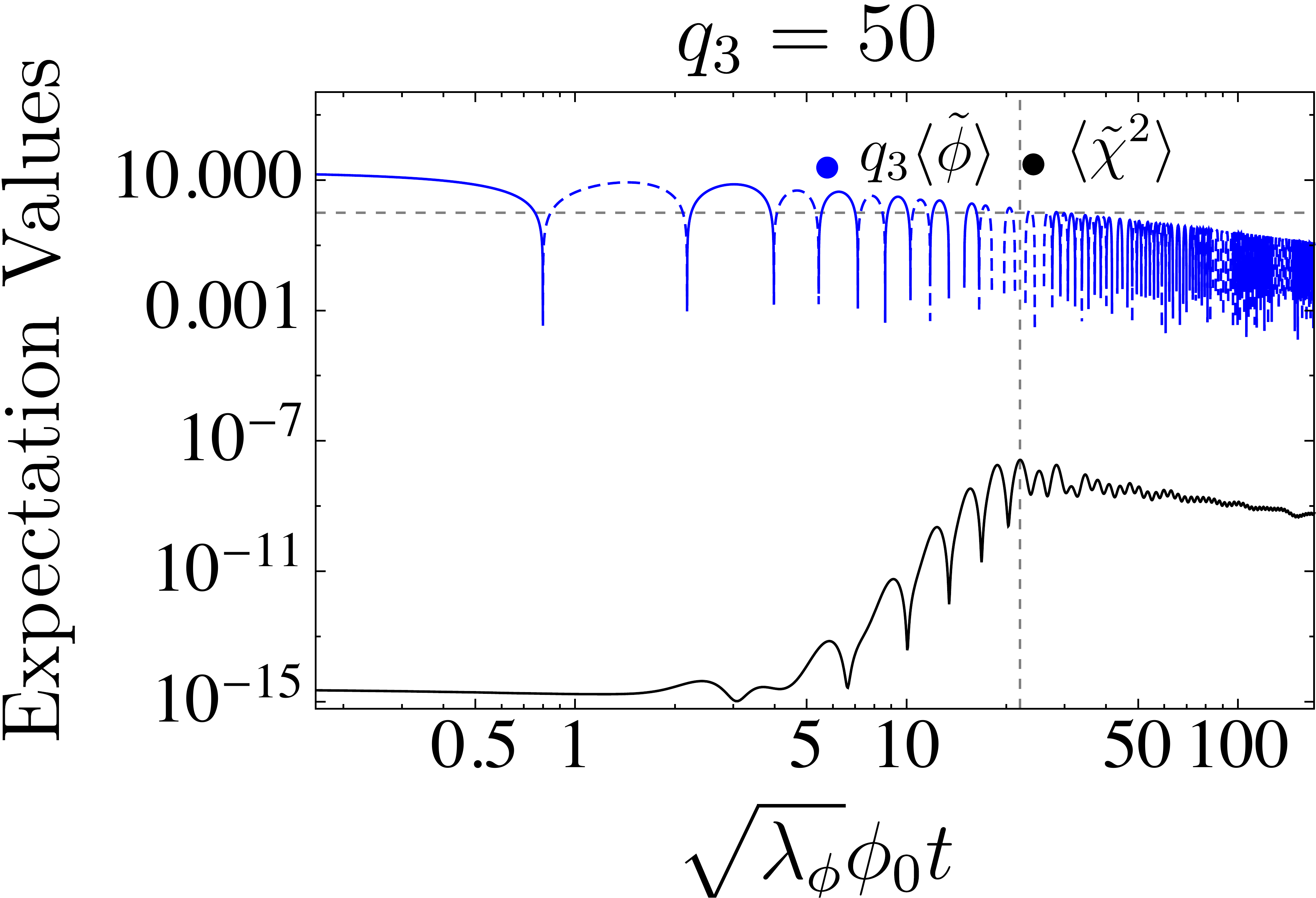}
    \includegraphics[width=0.45\textwidth,height=4.75cm]{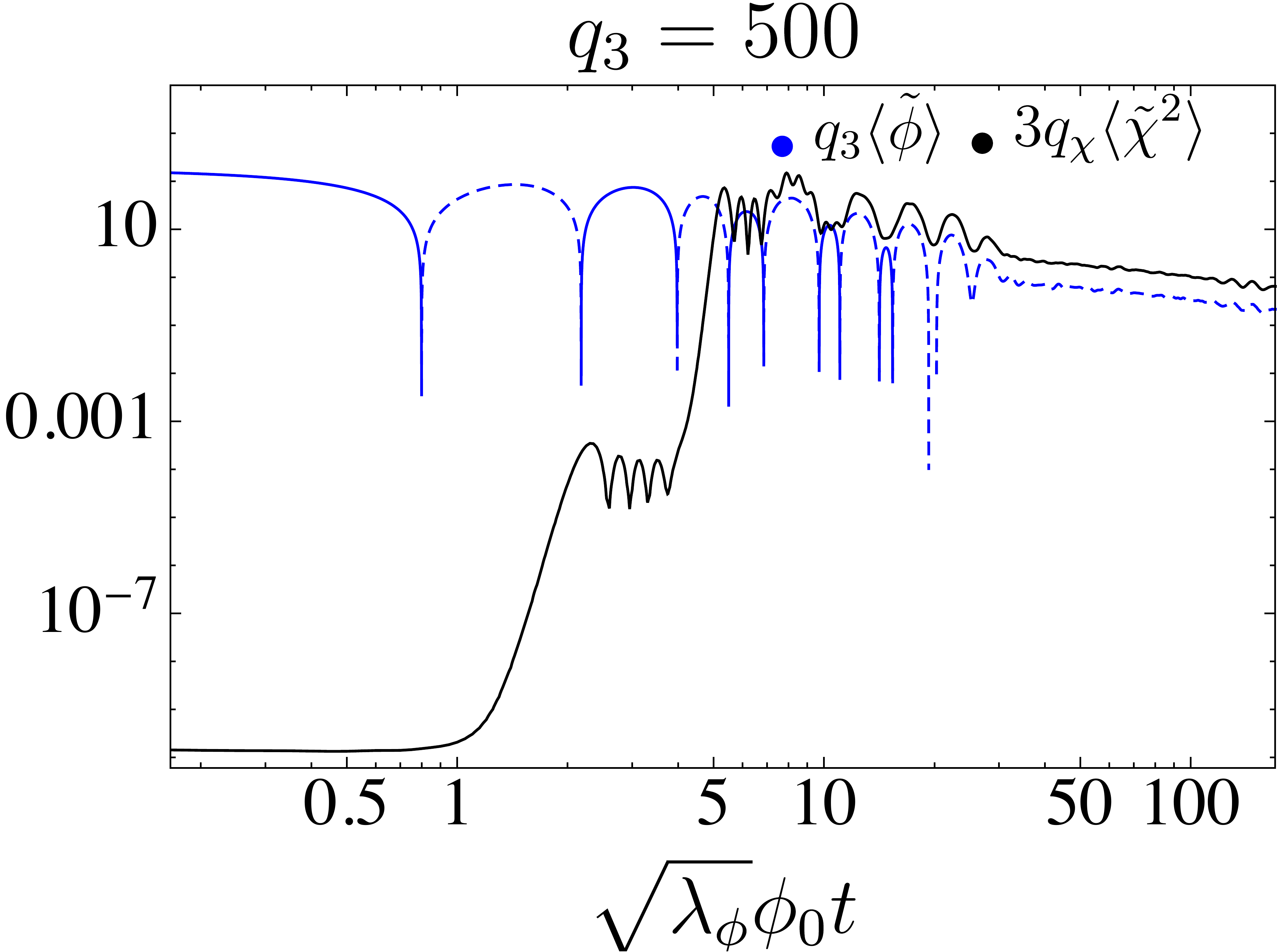} 
    \end{center}
    \vspace*{-0.5cm}
    \caption{{\it Left}: Time evolution of the inflaton's volume average $\langle\tilde{\phi}\rangle$ times the resonance parameter $q_{3}$ ($q_{3,\text{eff}}$), and the variance of the daughter field $\Varc$ for $q_{3}=50$. The horizontal dashed line represents $q_{3,\text{eff}}=1$, whereas the vertical dashed line is the time at which the maximum value of the variance, $\Varc_{max}$, is reached. {\it Right}: Time evolution of $q_{3} \langle \tilde{\phi}\rangle$ and the variance of the daughter field $\Varc$ times $3\,q_{\chi}$ for $q_{3}=500$. We see that once the term $3q_{\chi}\Varc$ overtakes $q_{3} \langle \tilde{\phi}\rangle$ the growth of the variance stops. For both panels, the solid (dashed) blue lines are the positive (negative) part of the inflaton mean value.} \label{fig:PhiVarchiP} 
\end{figure*}

The dynamics of preheating in the polynomial scenario can be described as follows. In an initial stage, the inflaton oscillates around its minimum, like in the monomial scenario, except that the nature of the potential of Eq.~\eqref{eq:inflatonpoly} makes the first oscillations to be an-harmonic. As the amplitude decays with the expansion of the universe, the oscillations become gradually more and more harmonic, as the potential becomes essentially more and more quadratic around the minimum. As within each oscillation the inflaton becomes negative during half of the oscillation, the effective mass $m_{\tilde{\chi}}$ of the $\tilde{\chi}_k$ modes becomes also negative for $\kappa < \sqrt{\sigma|\phi|}\,a$. This induces an exponential growth of such IR modes during the semi oscillations with $\phi < 0$, as can be clearly observed in Fig.~\ref{fig:PhiVarchiP}. As in the previous scenario we identify two regimes depending on the value of $q_{3}$:

$i)$ For $q_{3}\lesssim 100$, tachyonic resonance ends due to the expansion of the universe, since the effective resonance parameter $q_{3,\text{eff}}=q_{3}|\langle\tilde{\phi}\rangle|$ decreases with the scale factor. When $q_{3,\text{eff}} < 1$ is achieved, the system enters into a narrow regime where tachyonic resonance is not efficient anymore. This can be observed clearly in the {\it Left} panel of Fig.~\ref{fig:PhiVarchiP}.

$ii)$ For $q_{3} \gtrsim 100$, tachyonic resonance rather ends (before it becomes narrow) due to the presence of the self-interaction of the $\chi$ field. This happens when the $\chi$ self-interaction term in $m_{\tilde{\chi}}$ overtakes the trilinear interaction term. In that moment the variance $\left\langle \tilde{\chi}^{2}\right\rangle$ reaches a maximal critical value,  which can be expressed as
\begin{equation}
\left\langle \tilde{\chi}^{2}\right\rangle _{crit}\simeq\frac{q_{3}|\langle\tilde{\phi}\rangle|}{3\,q_{\chi}}\,\:.\label{eqn:Varcritpoly}
\end{equation}
The end of tachyonic resonance when $\left\langle \tilde{\chi}^{2}\right\rangle$ reaches its critical value (\ref{eqn:Varcritpoly}) can be observed in the {\it Right} panel of Fig.~\ref{fig:PhiVarchiP}.

\begin{figure*}[t] \begin{center}
    \vspace*{0.5cm}
    \includegraphics[width=0.55\textwidth,height=5.5cm]{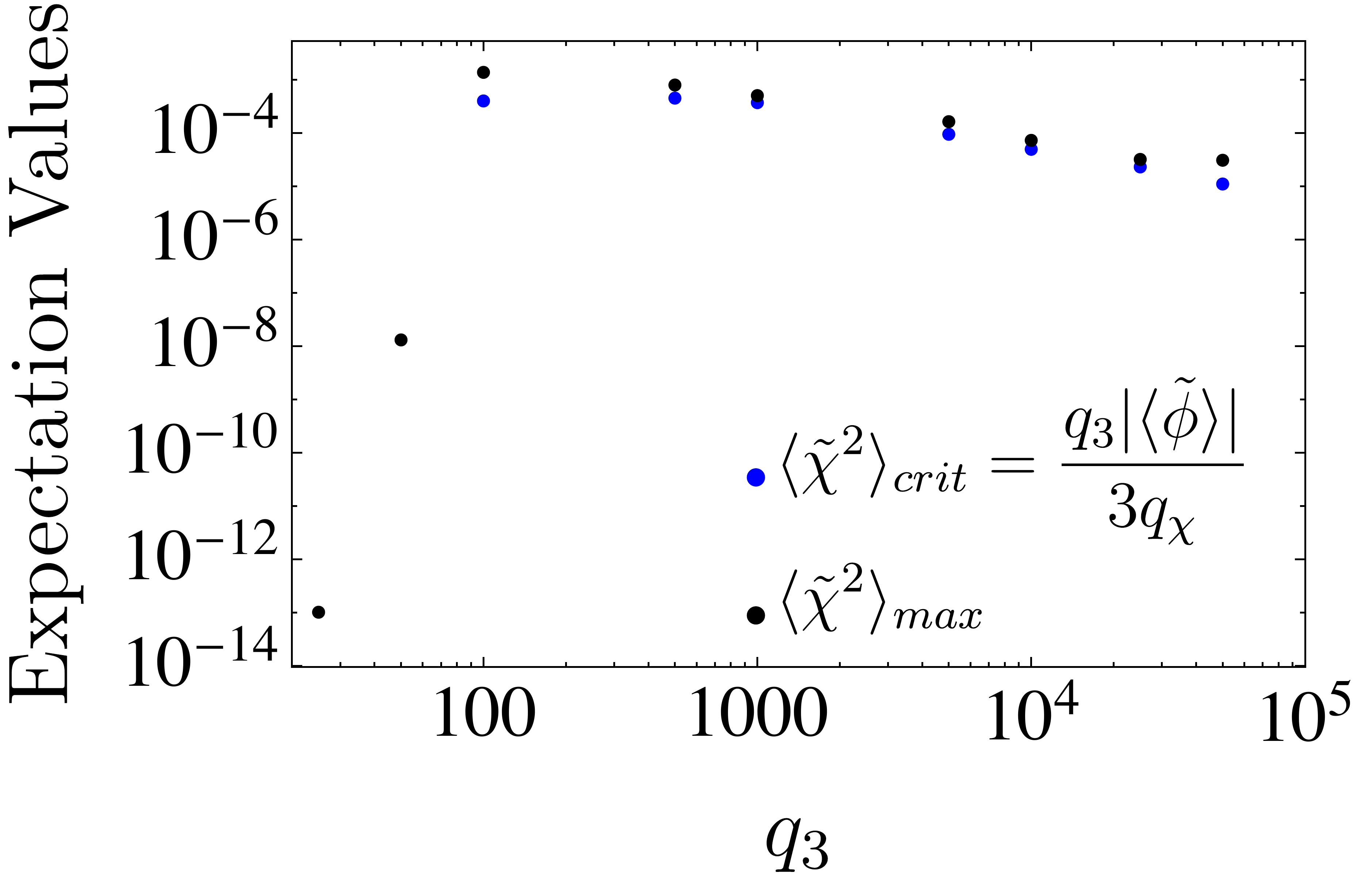} 
    \end{center}
    \vspace*{-0.5cm}
    \caption{Lattice results of $\Varc_{max}$ for different values of $q_3$ (black dots). It is clearly observed that for $q_3 \lesssim 100$ the amplitude of $\Varc_{max}$ is much smaller than for $q_3 \gtrsim 100$, as the tachyonic growth of $\Varc$ is simply stopped by the condition $q_{3,{\rm eff}} = 1$. We choose $\Varc_{max}$ as the value of $\Varc$ averaged over one oscillation when it reaches its maximum value. For comparison we also show the corresponding theoretical prediction for $q_3 \gtrsim 100$ in Eq.~\eqref{eqn:Varcritpoly} (blue dots).} \label{fig:Varchivsq3P}
\end{figure*}

In Fig.~\ref{fig:Varchivsq3P}, we show the theoretical prediction for $q_{3} \gtrsim 100$ of the maximum value of $\Varc$ given by Eq.~\eqref{eqn:Varcritpoly} (blue dots) and the actual values obtained in lattice simulations (black dots). We observe that for $q_{3}\lesssim100$ the variance $\Varc$ grows with $q_{3}$, while for $q_{3}\gtrsim100$ it decays inversely proportional to $q_{3}$, in correspondence with the theoretical description $\Varc_{crit} \simeq \frac{q_{3}\langle\tilde{\phi}\rangle}{3\, q_{\chi}} = \frac{8\langle\tilde{\phi}\rangle}{3\,q_{3}}$ (where in the last identity we have introduced the saturation relation $q_{\chi} = \frac{q_{3}^2}{8}$ from Eq.~(\ref{qchip q3p})). While our theoretical estimation in Eq.~(\ref{eqn:Varcritpoly}) captures only approximately the truly measured maximum variance $\Varc_{max}$, it describes nonetheless exactly its dependence on $q_{3}$. After the variance reaches the critical value $\langle\tilde{\chi}^{2}\rangle_{crit}$, in the following stages the system dynamics becomes fully non-linear due to the backreaction of the daughter field into the inflaton. Similarly as in the scenario with the monomial potential, due to the non-zero value acquired by the variance of $\chi$, the inflaton is forced to oscillate around a new minimum developed in its effective potential, a minimum that corresponds to a negative value of $\phi$.
\subsection{Gravitational wave production}\label{sec:GWsPoly}

\begin{figure}[t] \begin{center}
    \includegraphics[width=0.45\textwidth,height=5cm]{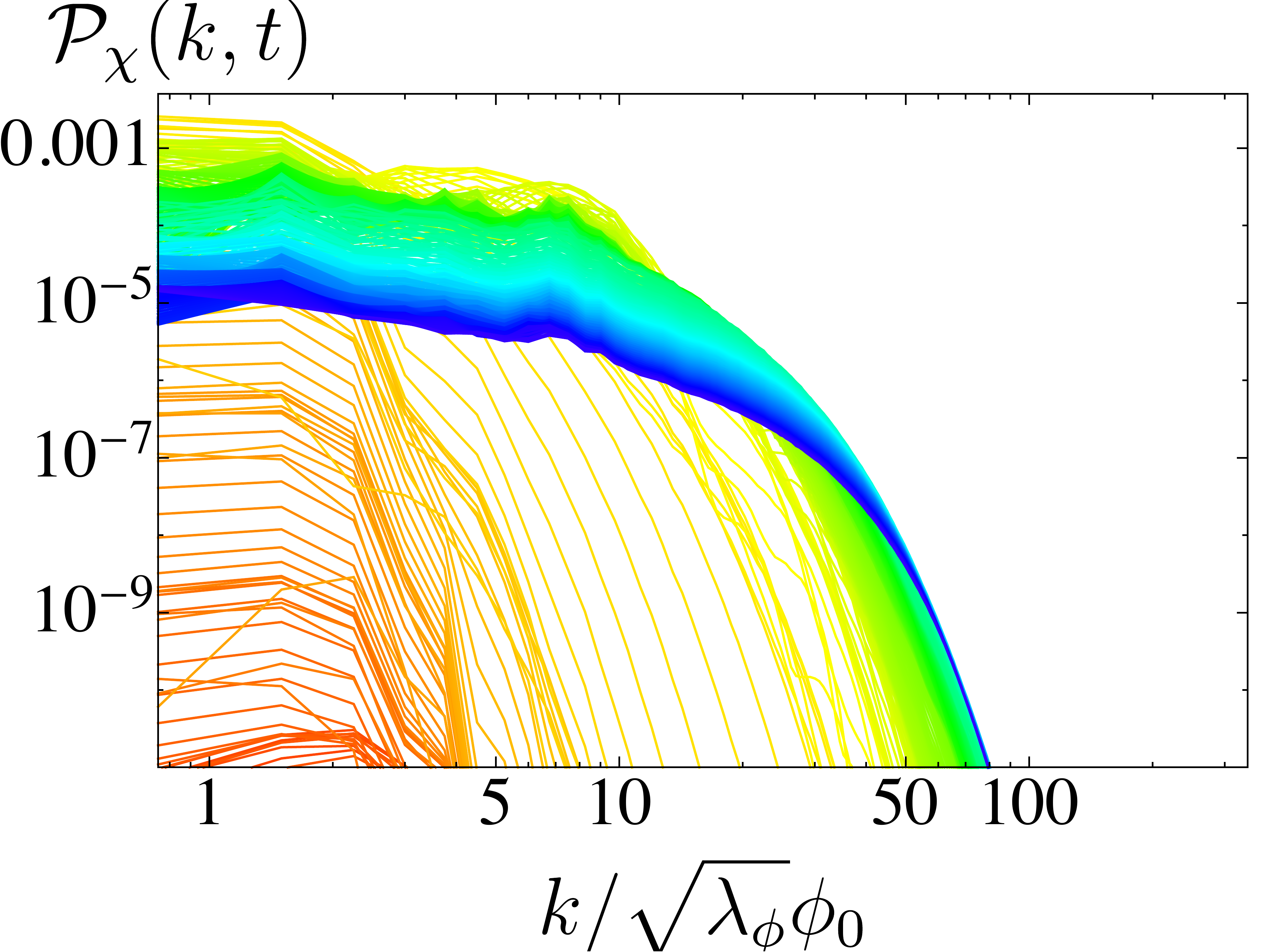} 
    \includegraphics[width=0.45\textwidth,height=5cm]{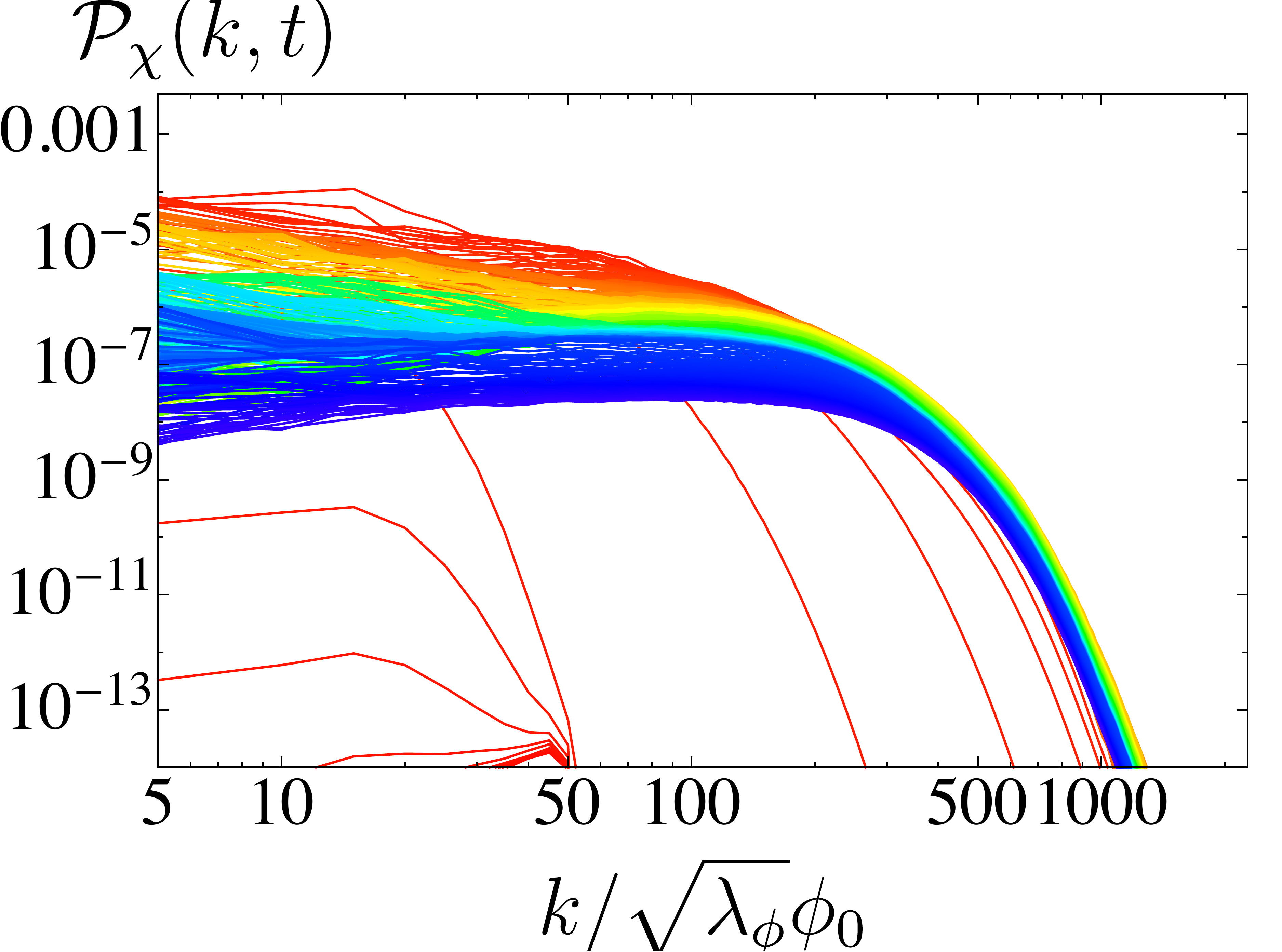} 
    \vspace*{0.5cm}
    \includegraphics[width=0.45\textwidth,height=5cm]{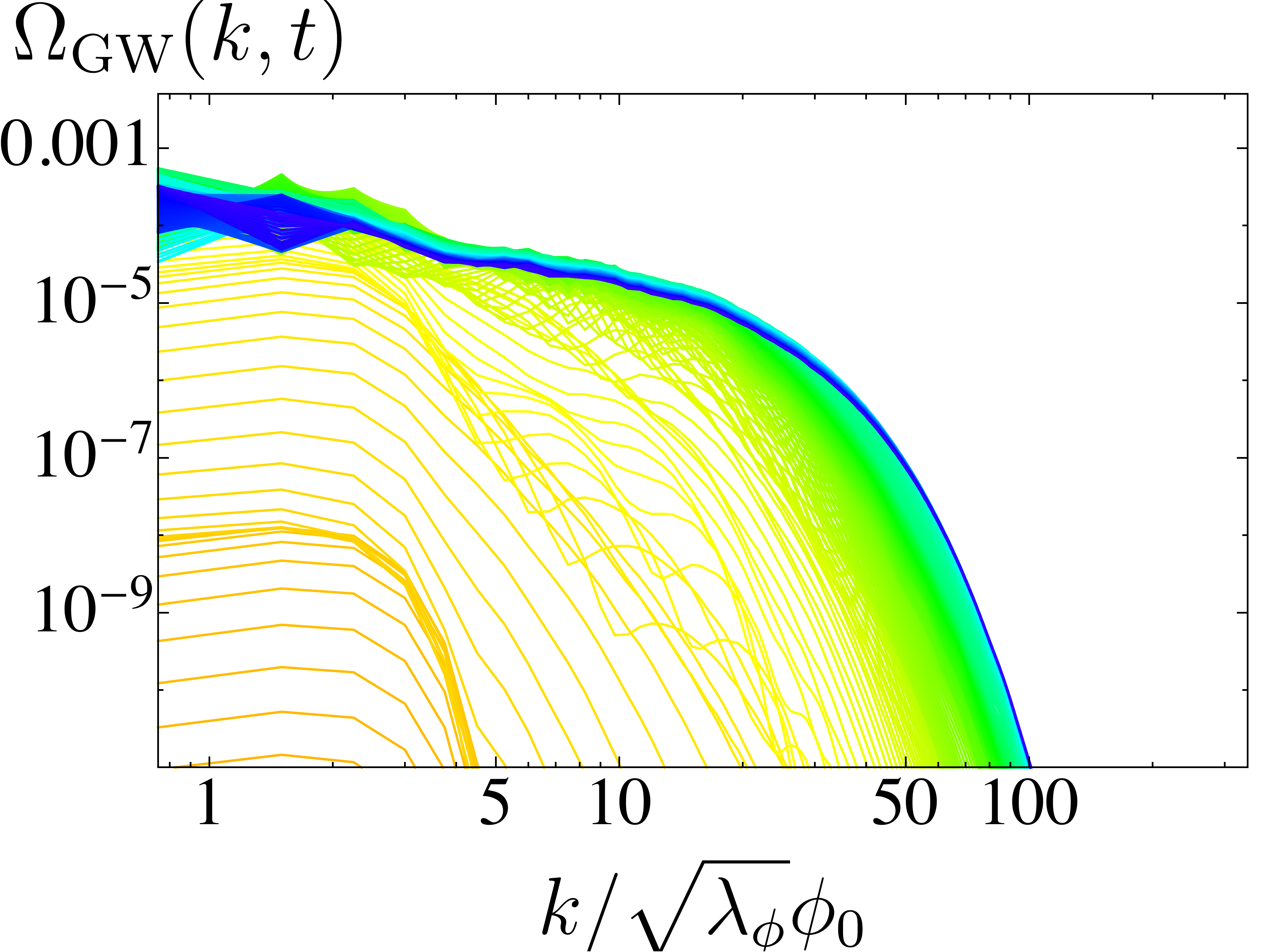} 
     \includegraphics[width=0.45\textwidth,height=5cm]{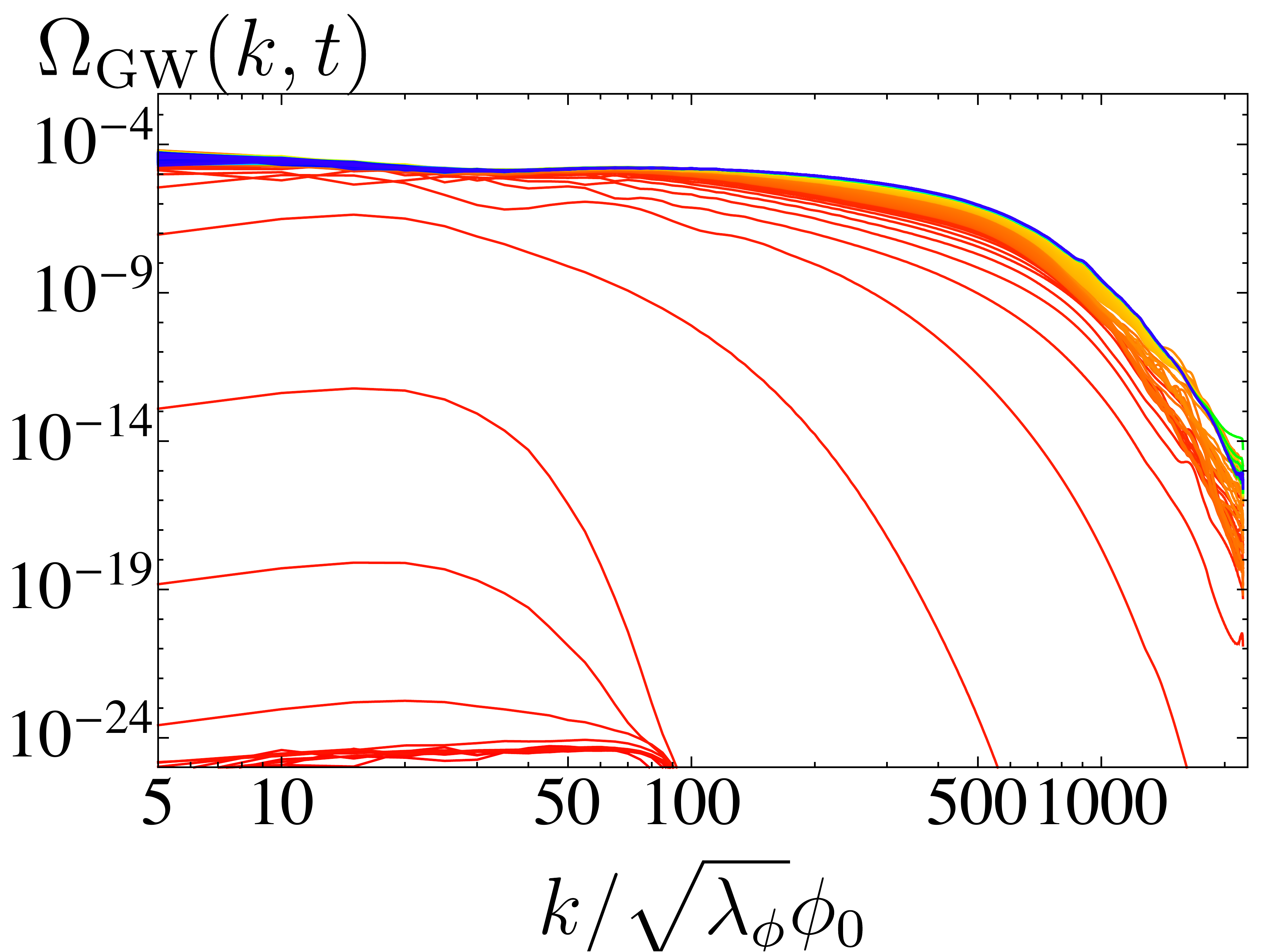} 
    \end{center}
    \vspace*{-0.5cm}
    \caption{{\it Top} panels: Matter power spectrum of the daughter field $\mathcal{P}_{\tilde\chi}(k,t)$, as a function of the momentum $k/m_{\phi}$ for $q_{3}=100$ ({\it Left}) and for $q_{3}=10^4$ ({\it Right}). {\it Bottom} panels: GW energy density power spectrum $\Omega_{\rm GW}(k,t)$, for $q_{3}=100$ ({\it Left}) and for $q_{3}=10^4$ ({\it Right}). In all panels, colors run from red in early times to blue in late times. Spectra are measured every $\Delta(\sqrt{\lambda_{\phi}}\phi_0 t)=0.2$ until saturation.} \label{figChiGWspectrumPoly} 
\end{figure}

In this section, we study the production of GWs during preheating after polynomial inflation. We follow the same procedure explained in Sec.~\ref{sec:GWsM2Phi2}, tracking the real time dynamics of the GWs and measuring their energy density power spectrum $\Omega_{\rm GW}(k,t)$, c.f.~Eq~\eqref{eqn:OmegaGW}. In Fig.~\ref{figChiGWspectrumPoly}, we show $\Omega_{\rm GW}(k,t)$ and the matter power spectrum of the daughter field $\mathcal{P}_{\tilde\chi}(k)$. Spectra are measured every $\Delta(\sqrt{\lambda_{\phi}}\phi_0 t) = 0.2$ intervals, up to a final time $\sqrt{\lambda_{\phi}}\phi_0 t_{\rm f} = 400$. We see that GWs are produced in pulses coinciding with the growth of the variance $\Varc$ of the daughter field. The production of GW stops around the time when $\Varc$ reaches its maximum critical value $\Varc_{crit}$. At this moment, the growth of $\Omega_{\rm GW}(k,t)$ ceases and the non-linearities developed in the dynamics of the inflaton-daughter system leads to a transfer of power into more ultraviolet (UV) modes in the GW spectrum.

\begin{figure}[t] \begin{center}
 \includegraphics[width=0.475\textwidth,height=4.5cm]{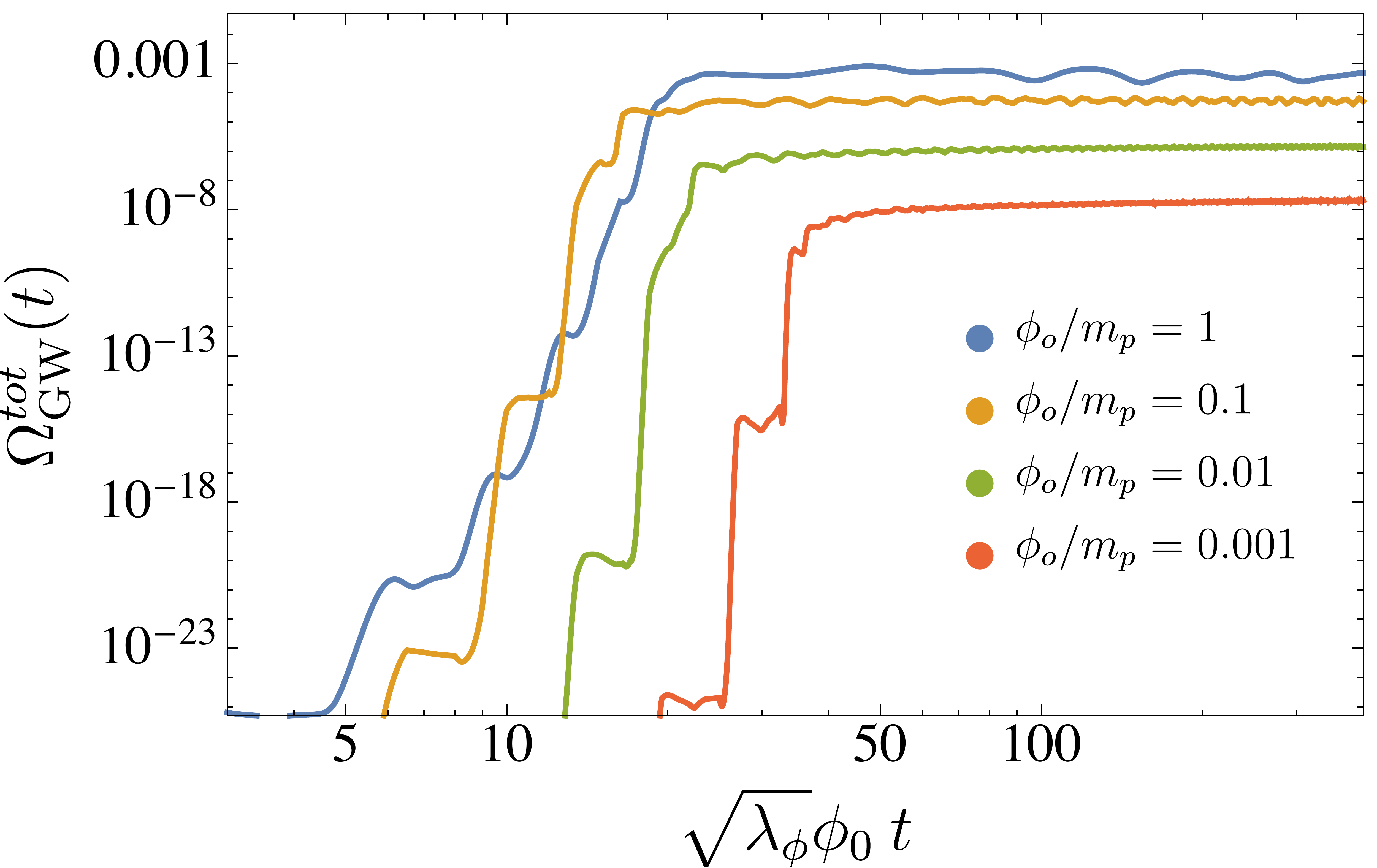} 
  \includegraphics[width=0.475\textwidth,height=4.5cm]{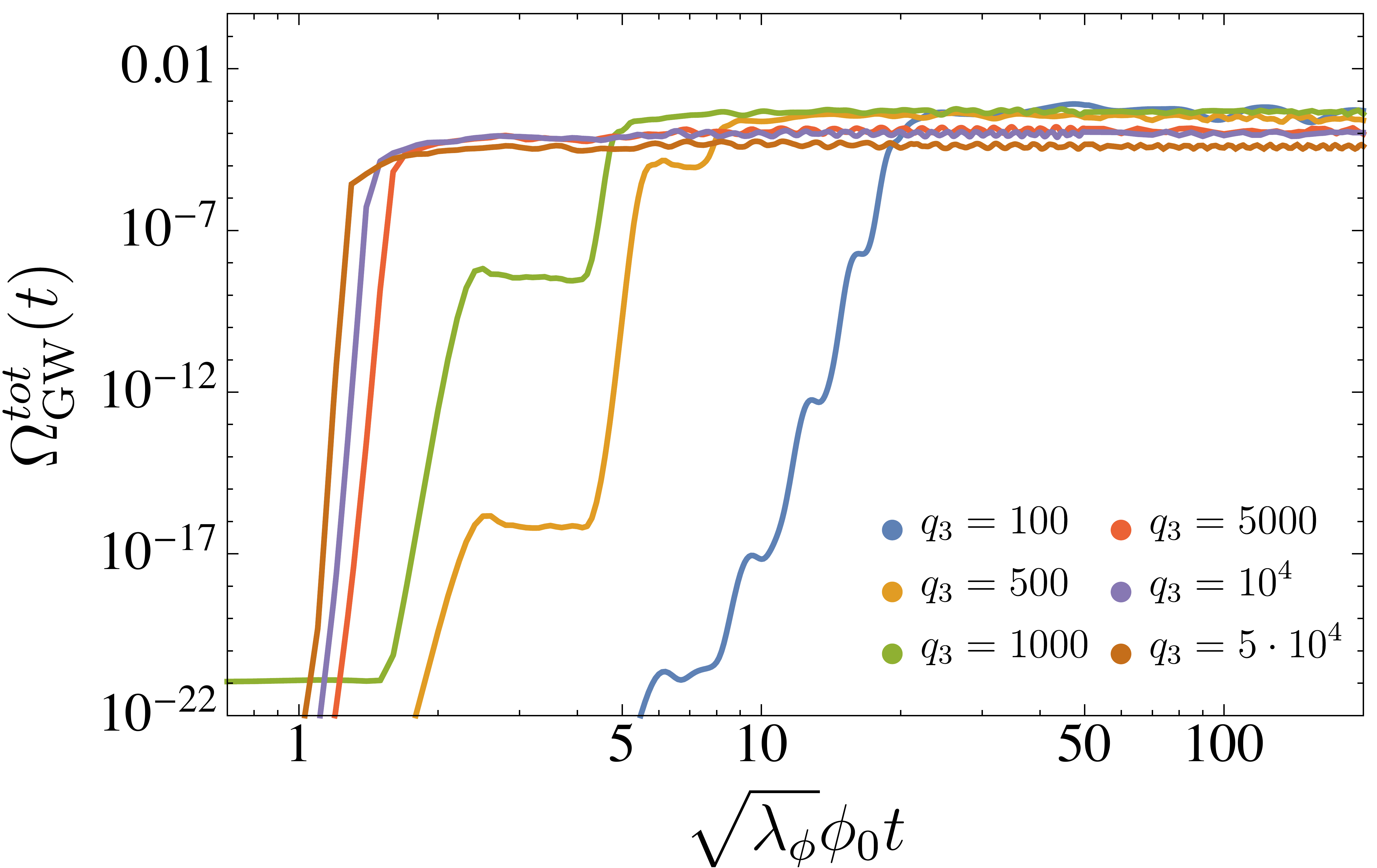}
    \end{center}
    \vspace*{-0.5cm}
    \caption{Time evolution of GWs total energy density: {\it Left:} for different values of $\phi_0$, with $q_{3} = 100$ and $q_{\chi}=q_{3}^2/8$. We observe that the largest amount of GWs is produced by $\phi_0=m_p$. {\it Right:} For different values of $q_{3}$, with $q_{\chi}=q_{3}^2/8$ and $\phi_0 = m_p$.}
    \label{fig:rhopolyphi0q3p}
\end{figure}

Fixing $q_{\chi} = q_{3}^2/8$ according to the saturation relation, we have characterized the amplitude and the peak position of the GW background as a function of $\phi_0$ and $q_{3}$. In the {\it Left} panel of Fig.~\ref{fig:rhopolyphi0q3p}, we plot the time evolution of the GW total energy density for different values of $\phi_{0}$, and observe that the total GW energy density decreases with $\phi_0$. On the other hand, decreasing $\phi_0$ implies decreasing the energy scale of inflation and hence also the peak frequency of the signal today $f_{\rm GW}^{(p,0)}$. In the {\it Right} panel of Fig.~\ref{fig:rhopolyphi0q3p}, we plot the time evolution of the GW total energy density for different values of $q_{3}$, and observe that the final amplitude decreases mildly when increasing this parameter. At the same time, the dependence of $f_{\rm GW}^{(p,0)}$ on the resonance parameter $q_{3}$ is determined by the initial instability band, $\kappa < \sqrt{q_{3}|\tilde{\phi}|}\,a$. Regarding the expansion history of the universe, which is relevant for the determination the $\epsilon_{\rm f}$ parameter to redshift correctly the signal, our lattice simulations indicate that for sufficiently large couplings $q_{3} \gtrsim 10^3$ the energy budget of the system is dominated by kinetic and gradient energy of the daughter field, which behaves as radiation. The universe is therefore in RD when we end our simulations, and therefore we take $\epsilon_{\rm f} = 1$. Using Eqs.~\eqref{eqn:redshiftfreq} and~\eqref{eqn:redshiftamp}, we obtain a parametrization of the SGWB peak position and amplitude as function of $\phi_0$ and $q_{3}$, valid for $q_{3} \gtrsim 10^3$ and $3\times10^{-5}m_p\leq\phi_0\leq m_p$, as
\begin{eqnarray}
    f_{\rm GW}^{(p,0)}(\phi_0,q_{3}) &\simeq& (5.05 \pm 2.03)\times10^6\: \left(\frac{\phi_0}{m_p}\right)^{0.31 \pm 0.1}\:\left(\frac{q_{3}}{100}\right)^{0.4\pm0.06}  \: \text{Hz}\: ,\\
    h_0^2\Omega^{(p,0)}_{\rm GW}(\phi_0,q_{3}) &\simeq& (5.0 \pm 1.5)\times 10^{-9}\:
    \left(\frac{\phi_0}{m_p}\right)^{1.41 \pm 0.13}\:\left(\frac{q_{3}}{100}\right)^{-0.76 \pm 0.06}\:.
\end{eqnarray}\label{eqn:fGWAmpPoly}
From the parametrization we observe that the maximum amplitude for the SGWB is obtained for $\phi_0 \simeq m_p$, but in that case the background is peaked at very high frequencies $f_{\rm GW}^{(p,0)} \gtrsim 10^7$ Hz. While decreasing $\phi_0$ allow us to move the frequency window to smaller values, down to values $f_{\rm GW}(\phi_0 = 10^{-5}\:m_p)\sim10^{5}$ Hz, this can only be done the expense of suppressing the amplitude of the SGWB spectrum, which then becomes $h_o^2\Omega^{(0)}_{\rm GW}(\phi_0 = 10^{-5}\:m_p) \lesssim 10^{-15}$. Furthermore, the dependence on the resonance parameter $q_{3}$ just follows a similar behavior as other in parametric excitation mechanism~\cite{Figueroa:2017vfa}, so that increasing $q_{3}$ parametrically shifts the signal to higher frequencies while decreasing its amplitude. 

\begin{figure}[t] 
\begin{center}
\vspace*{0.5cm}
    \includegraphics[width=0.55\textwidth,height=5.5cm]{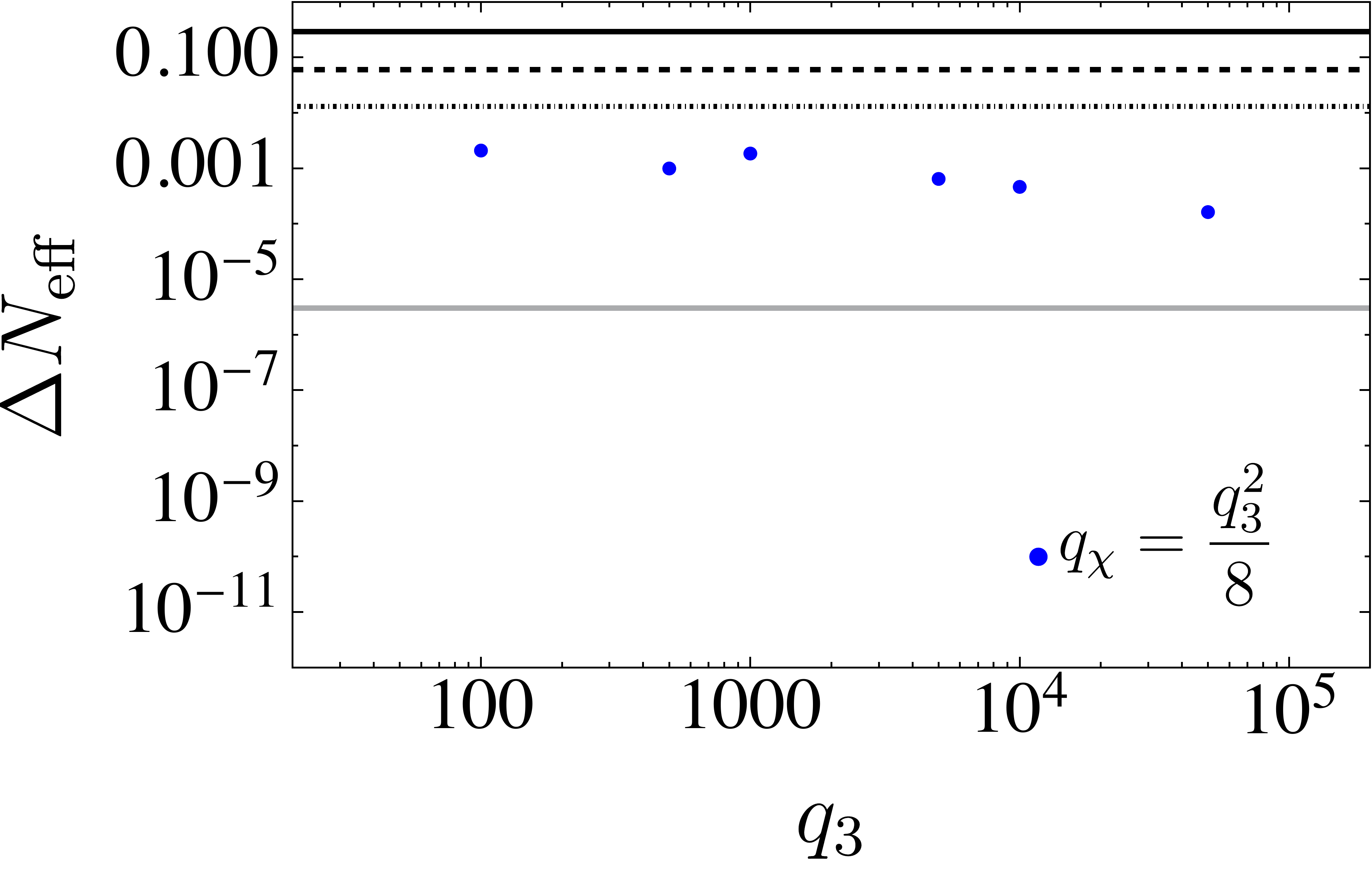} \hspace{0.5cm}
    \end{center}
    \vspace*{-0.5cm}
    \caption{Number of relativistic degrees of freedom beyond Standard Model, $\Delta N_{\rm eff}$, for different values of $q_{3}$ with $q_{\chi}=q_{3}^2/8$. The top solid black line corresponds to the current bound coming from Planck data $\Delta N_{\rm eff}\lesssim 0.29$, the dashed line below represents the future CMB-S4 $\Delta N_{\rm eff} \lesssim 0.06$ upper bound at $2\sigma$. The dot-dashed black line corresponds the projected bounds at $2\sigma$ on $\Delta N_{\rm eff}\lesssim 0.013$ from next-generation satellite missions, whereas the bottom gray solid line corresponds to a hypothetical cosmic variance limited (CVL) CMB polarization experiment upper bound  $\Delta N_{\rm eff}\lesssim 3\cdot10^{-6}$.} 
    \label{fig:Neffpoly}
\end{figure}

In conclusion, despite the fact that in the polynomial modeling we achieve partially our intention to shift towards smaller frequencies the SGWB, this can only be done at the expense of suppressing the amplitude of the signal. In order to have a large amplitude background we are forced to have it peaked, as usual, at very high frequencies, where direct detection experiments are yet far from the desired capabilities to detect a background like this one.
%
%
Given the inability to detect the SGWB directly also in this model, we can still see whether constraints on a maximum trilinear interaction coupling could be placed using current and planned constraints on $\Delta N_{\rm eff}$. Recalling  Eq.~\eqref{eqn:Neff}, we translate the energy density of the GW background into $\Delta N_{\rm eff}$ and compare with the present/projected bounds on $\Delta N_{\rm eff}$, as we can see in Fig.~\ref{fig:Neffpoly}. We find that for $q_{3} = 100$ and $\phi_0 = m_p$, we obtain a maximum value as $\Delta N_{\rm eff}\simeq 0.001$, which is well above a futuristic CVL experiment, but still below the more realistic expected bounds in near future experiments. We conclude therefore that it is not possible to probe directly or indirectly the predicted signal with upcoming experiments.

\section{Discussion}\label{sec:Discussion}
In this work, we have studied the GW production during
preheating due to a trilinear interaction $\phi\chi^2$ between the inflaton $\phi$ and a daughter field $\chi$. Considering first a preheating scenario with the inflaton having a quadratic monomial potential $\frac{1}{2}m_{\phi}^{2}\phi^{2}$, we added a self-interaction term for the daughter field $\frac{1}{4}\lambda\chi^{4}$,
to ensure the total potential to be bounded from below. This imposes a relation between
the resonance parameters $q_{3}$ and $q_{\chi}$, which control the
intensity of the daughter field excitation. Only values for which the relation
$q_{\chi}>\frac{q_{3}^{2}}{2}$ is satisfied avoid runaway solutions of $\chi$. Using \CL~to perform lattice simulations of the system, we find that for $q_{3}<10$ the growth of
the daughter field variance $\left\langle \tilde{\chi}^{2}\right\rangle$ stops at an early moment due to the expansion of the universe, determined by when the system enters into narrow resonance. For $q_{3}>10$ instead, the variance grows up to a critical value $\langle{\tilde{\chi}^{2}}\rangle_{crit}$, determined by when the $\chi$ self-interaction overtakes the trilinear interaction. GW production follows the step-like growth of the
daughter field power spectrum, and ceases once the variance of $\chi$ reaches its critical value $\langle{\tilde{\chi}^{2}}\rangle_{crit}$, see Figs.~\ref{fig:PhiVarchi} and \ref{fig:rhoGW}.
While the SGWB produced in this preheating scenario can reach an amplitude as large as
$h_{0}^{2}\Omega_{{\rm GW}}^{(0)}\simeq10^{-9}$, the signal is peaked at
high frequencies $f_{{\rm GW}}\gtrsim 10^{8}$ Hz, which precludes GW direct detection experiments to probe this background. A large amount of GWs is generated nevertheless in this scenario, which can yield a value in terms of an effective number of relativistic species as large as
$\Delta N_{{\rm eff}}\sim0.003$ (corresponding to $q_{3}=9$). Unfortunately, this is still below the projected sensitivity on $\Delta N_{{\rm eff}}$ of next-generation CMB experiments.

We have also studied the case where the inflaton has a polynomial
potential (both during preheating and inflation), whose shape is controlled by a single parameter, the value of a near inflection point $\phi_{0}$ in the potential, which can range from $3\times10^{-5}m_{p}<\phi_{0}<m_{p}$. This scenario might improve the situation in two directions: a) decreasing the energy scale of inflation, hence shifting to smaller
frequencies the SGWB peak; b) increasing the amount of GW energy density produced relative to the field energy budget of the system. In this scenario, runaway solutions of $\chi$ are avoided if the resonance parameters satisfy the relation $q_{\chi}>\frac{q_{3}^{2}}{8}$. Using \CL~to perform lattice simulations of the system, we find that for $q_{3}\lesssim100$, tachyonic resonance ends soon after the onset of inflaton oscillations, when the system enters into narrow resonance due to the expansion of the universe. For $\ensuremath{q_{3}\gtrsim100}$ instead, tachyonic resonance stops when the variance of the daughter field reaches a critical value $\left\langle \tilde{\chi}^{2}\right\rangle _{crit}$, which occurs
when the $\chi$ self-interaction term overtakes
the trilinear interaction term. Regarding GW production, we find that choosing the highest possible value of the  inflection point, $\phi_{0}=m_{p}$,
leads to the largest GW production, with a SGWB amplitude as large as $h_{0}^{2}\Omega_{{\rm GW}}^{(0)}\simeq5\cdot10^{-9}$. This signal is however still peaked at large $f_{{\rm GW}}\simeq5\cdot10^{6}$ Hz, which cannot be probed by planned GW experiments. While we can shift the signal down to $\sim 10^{5}$ Hz frequencies by decreasing $\phi_0$ down to its minimum value compatible with CMB constraints, we can only do it at the expense of suppressing the signal amplitude down to $h^2\Omega_{\rm GW}^{(p,0)} \simeq 10^{-15}$. In terms of radiation degrees of freedom represented by the SGWB in this model, we find a maximum value as $\Delta N_{{\rm eff}}\simeq0.001$, which is still out of the reach of near future next-generation CMB experiments. 

In summary, preheating scenarios with an inflaton monomial or polynomial potential and a trilinear coupling between the inflaton a daughter field, can lead to very efficient GW production, but the resulting GW signal is still out of the reach of current and planned direct and indirect probes of SGWBs. One direction we may explore in the future is to consider the same type of coupling in lower scale inflationary scenarios, for which the SGWB frequency peak should be shifted to smaller values, while there is no particular reason to expect that the efficiency of GW production should be decreased.

\acknowledgments 

We thank Djuna Croon for interesting discussions and collaboration at early stages of the project. Our work is supported by the Generalitat Valenciana grant PROMETEO/2021/083, and by the Spanish Ministerio de Ciencia e Innovacion grant PID2020-113644GB-I00. C.C.~is also supported by the grant PROMETEO-2019-083, the European Union’s Horizon 2020 research and innovation programme under the Marie Sk\l{}odowska-Curie grant agreement No 860881-HIDDeN, and partially by the FCT project Grant No.~CERN/FIS-PAR/0027/2021.

\appendix
\section{Gravitational wave evolution in the lattice}\label{app:GWs}
In order to evolve the $h_{ij}$ fields in the lattice, we follow the method introduced in~\cite{Garcia-Bellido:2007fiu}. We evolve a set of fields $u_{ij}$ with the equation of motion
\begin{equation}\label{eqn:usEoM}
    \ddot{u}_{ij}+3H\dot{u}_{ij} - \dfrac{\nabla^2}{a^2} u_{ij} = \dfrac{2}{m_p^2 a^2}\{\partial_i \phi \partial_j \phi + \partial_i \chi \partial_j \chi\} \:.
\end{equation}
The physical $h_{ij}$ tensor perturbations are obtained through
\begin{equation}\label{eqn:TTprojector1}
    h_{ij}(k,t)= \Lambda_{ijkl}(\hat{k})\: u_{kl}(k,t) \:,
\end{equation}
where $\Lambda_{ijkl}$ is defined by
\begin{eqnarray}
\label{eqn:ProjectorTT2}
    \Lambda_{ijlm}({\bf \hat{k}}) \equiv P_{il}({\bf \hat{k}})  P_{jm}({\bf \hat{k}}) - \dfrac12 P_{ij}({\bf \hat{k}}) P_{lm}({\bf \hat{k}}) \: ,\\
     P_{ij}= \delta_{ij} - \hat{k}_i \hat{k}_j\:,~~  \hat{k}_i = k_i/k \:. ~~~~~~~~~
\end{eqnarray}
The method relies on the fact that at any time we need to compute the GW energy density power spectrum, we fourier transform $u_{ij}$ fields and then project them using Eq.~\eqref{eqn:TTprojector1}. This is possible because both operations commute, therefore it is not necessary to have the source projected to TT space at every time step of evolution, which would take a very long time for the simulations as the TT projection is done in Fourier space and one would need to go back and forth between real and Fourier space.

In the lattice, the TT projection is done with an equivalent lattice projector $\Lambda_{ij,lm}^{\rm (L)}$ written in terms of a lattice momentum $k_i^{\rm (L)}$ which depends on the choice of spatial-derivative, see~\cite{Figueroa:2011ye} for a discussion. We use the nearest-neighbor derivative of equation (71) in~\cite{Figueroa:2020rrl}, for which the lattice momenta is given by 
\begin{equation}
    k^{(L)}_{i} = 2 \dfrac{\sin (\pi \tilde{n}_i/N)}{dx} \: .
\end{equation}

\section{Frequency and amplitude of the SGWB today}\label{app:Redshift}
The Stochastic Gravitational Wave Background (SGWB) today is obtained by redshifting the amplitude and frequency of the background from the end of GWs production, $t_{end}$,
following the appropriate expansion history of the Universe. We know the expansion history up to the final time $t_{\rm f}$ of our simulations, and noticing that $t_{end} < t_{\rm f}$, we can redshift from $t_{\rm f}$ until the present time $t_0$. Let us characterize the expansion history between $t_{\rm f}$ and the onset of radiation domination (RD) at $t_{\rm RD}$, with an effective equation of state $\bar{w}=p/\rho$
\begin{eqnarray}
    \log\left(\dfrac{\rho^{\rm (RD)}_{tot}}{\rho^{\rm (f)}_{tot}}\right) &=& -3 \int^{a_{\rm RD}}_{a_{\rm f}} \dfrac{da'}{a'}(1+w(a')) \: ,  \\
     &=& -3(1+\bar{w}) \log\left(\dfrac{a_{\rm RD}}{a_{\rm f}}\right) \: ,
\end{eqnarray}
The SGWB spectrum actually peaks at some sub-horizon scale $k_p = \beta_p a_{\rm f} H_{\rm f} $, with $a_{\rm f} = a(t_{\rm f})$ and $H_{\rm f }= H(t_{\rm f})$. We expect the SGWB spectrum to peak therefore at a frequency $f_{\rm GW}$ today
\begin{eqnarray}
f_{\rm GW} &\equiv& \dfrac{k_{\rm GW}}{2\pi a_0} = \dfrac{\beta_{\rm GW}}{2\pi} \left(\dfrac{a_{\rm f}}{a_{\rm RD}}\right)\left(\dfrac{a_{\rm RD}}{a_0}\right) H_{\rm f} \nonumber\\
     &=& \dfrac{k_p}{a_{\rm f}H_{\rm f}}  \epsilon_{\rm f}^{1/4}G_{\rm RD}^{1/4}\left(\dfrac{H_{\rm f}}{m_p}\right)^{1/2} \left(\dfrac{2}{3}\right)^{1/4}  \dfrac{\rho_{\rm rad,0}^{1/4}}{2 \pi} \, ,
\end{eqnarray}
where we have defined 
\begin{equation}
\hspace*{-0.4cm}\epsilon_{\rm f} \equiv \left(\dfrac{a_{\rm f}}{a_{\rm RD}}\right)^{1-3\bar{w}}\,,~~ G_{\rm RD} \equiv \left(\dfrac{g_{\rm RD}}{g_{0}}\right)\left(\dfrac{g_{s,0}}{g_{s,RD}}\right)^{4/3}\hspace*{-0.1cm},
\end{equation}
with $g_{s,t}$ and $g_{t}$ the entropic and energy density relativistic degrees of freedom. We can characterize the factor $G_{\rm RD}^{1/4}$ by taking into account that the SM degrees of freedom above the electroweak scale amount to $g_{s,RD}=g_{\rm RD}=106.75$, so we write $G_{\rm RD}^{1/4}\simeq (g_{s,RD}/100)^{-1/12}$. The redshift factor finally reads, using $\rho_{rad,0} = 3.37\times10^{-51}\text{GeV}^4$, 
\begin{equation}
    f_{\rm GW} \simeq 4\times 10^{10} \, \epsilon_{\rm f}^{1\over4} \left(\dfrac{g_{\rm s,RD}}{100}\right)^{-{1\over12}} \: \dfrac{k}{a_{\rm f} H_{\rm f}} \left(\dfrac{H_{\rm f}}{m_p}\right)^{1\over2} \: \text{Hz} \: .
\end{equation}
The redshifted GW spectrum amplitude is 
\begin{equation}
        h_0^2\Omega_{\rm GW}^{(0)}(f) = \epsilon_{\rm f} \:G_{\rm RD} \:h_{0}^2 \Omega_{rad}^{(0)} \:\Omega_{\rm GW}(k) \:, 
\end{equation}
and using the value of $h_{0}^2 \Omega_{\rm rad}^{(0)} \approx 4\times10^{-5}$, the peak amplitude of the spectrum today is
\begin{equation}
        h_0^2\Omega_{\rm GW}^{(0)}\Big|_{\rm peak} \simeq 1.6 \times 10^{-5} \: \epsilon_{\rm f} \: \left(\dfrac{g_{\rm s,RD}}{100}\right)^{-1/3} \: \Omega_{\rm GW}^{\rm (p)} \: .
\end{equation}

\footnotesize{
\bibliographystyle{JHEP}
\bibliography{refs}}

\end{document}